\documentclass[3p, twocolumn]{elsarticle}
\usepackage[hidelinks]{hyperref}

\journal{Journal of Network and Computer Applications}

\usepackage{lipsum}
\usepackage{amsmath}
\usepackage{latexsym}

\usepackage{tablefootnote}

\usepackage{soul}
\usepackage{color}
\usepackage{bm}
\usepackage{multirow}
\usepackage[anythingbreaks]{breakurl}
\usepackage[table,xcdraw]{xcolor}
\usepackage{array,booktabs}
\usepackage{balance}
\usepackage{setspace}

\usepackage{amssymb}
\usepackage{pifont}
\newcommand{\cmark}{\large\ding{51}}%
\newcommand{\xmark}{\large\ding{55}}%

\usepackage{tikz}
\newcommand*\circled[1]{\tikz[baseline=(char.base)]{
            \node[shape=circle,draw,inner sep=1pt,font=\sffamily\footnotesize] (char) {#1};}}
\usepackage{pifont}

\definecolor{lightgray}{gray}{0.9}
\definecolor{LightCyan}{rgb}{0.88,1,1}

\newcommand{\lmttfont}{\fontfamily{lmtt}\selectfont}

\newcommand{\revision}[1]{{\textcolor{black} {#1}}}

\newcolumntype{L}[1]{>{\raggedright\arraybackslash}m{#1}}
\newcolumntype{C}[1]{>{\centering\arraybackslash}m{#1}}
\newcolumntype{R}[1]{>{\raggedleft\arraybackslash}m{#1}}
\newcommand{\toolname}[1]{\emph{ThorFI}}
\newcommand{\netcastAgent}[1]{front-end}
\newcommand{\injectorAgent}[1]{injection agent}

\usepackage{amsmath}
\usepackage{bm}

\usepackage{caption}
\usepackage{subcaption}
\usepackage[flushleft]{threeparttable}

\begin{document}

\begin{frontmatter}

\title{ThorFI: A Novel Approach for\\Network Fault Injection as a Service}

\author{Domenico Cotroneo}
\ead{cotroneo@unina.it}
\author{Luigi De Simone\corref{correspondingauthor}}
\ead{luigi.desimone@unina.it}
\author{Roberto Natella}
\ead{roberto.natella@unina.it}

\address{DIETI - Università degli Studi di Napoli Federico II, Via Claudio 21, 80125 Napoli, Italy}

\begin{abstract}
In this work, we present a novel fault injection solution (\toolname{}) for virtual networks in cloud computing infrastructures. \toolname{} is designed \revision{to provide non-intrusive fault injection capabilities for a cloud tenant, and to isolate injections from interfering with other tenants on the infrastructure.} We present the solution in the context of the OpenStack cloud management platform, and release this implementation as open-source software. \revision{Finally, we present two relevant case studies of \toolname{}, respectively in an NFV IMS and of a high-availability cloud application. The case studies show that \toolname{} can enhance functional tests with fault injection, as in 4\%-34\% of the test cases the IMS is unable to handle faults; and that despite redundancy in virtual networks, faults in one virtual network segment can propagate to other segments, and can affect the throughput and response time of the cloud application as a whole, by about 3 times in the worst case.}
\end{abstract}

\begin{keyword}
Network Fault Injection; Virtualization; Chaos Engineering; OpenStack.
\end{keyword}

\end{frontmatter}


\section{Introduction}

Modern services for critical domains, including telecommunications, healthcare, finance, and more, run on \emph{virtual networks} that are deployed over cloud data centers. The emerging \emph{Network Function Virtualization} paradigm is a relevant example, as it is turning traditional network equipment and services (e.g., IMS, DPS, EPC, etc.) into software running on the cloud \cite{NFVISG2012a}. These software applications demand ultra-high reliability (e.g., \emph{five nines}) and responsiveness (e.g., can only tolerate few \emph{ms} of latency) \cite{TL9000, NFV-REL001, virtualization_survey}.

However, the virtualization of network functions raises performance and reliability concerns, due to faults occurring in the hardware, software, and configuration of cloud data centers \cite{gill2011understanding,lu2013cloud,cotroneo2017nfv,di2018availability,martini2019experimenting}. The frequency of these faults is high, due to the \emph{large scale} (up to tens of thousands of servers and network elements) and \emph{complexity} of data centers (e.g., due to sophisticated software-defined technologies) \cite{gunawi2014bugs, gunawi2016does, gunawi2018fail}. The problem is exacerbated by the massive adoption of \emph{off-the-shelf hardware and software components}. Examples of recurring faults are silent packet drops (e.g., due to faulty switches, where discard counters are zero), loops and load imbalance (e.g., buggy flow control algorithms and bad capacity planning), and proprietary protocol bugs (e.g., incompatibilities with other equipment).

\begin{figure*}[!h]
  \centering
  \includegraphics[width=1.8\columnwidth]{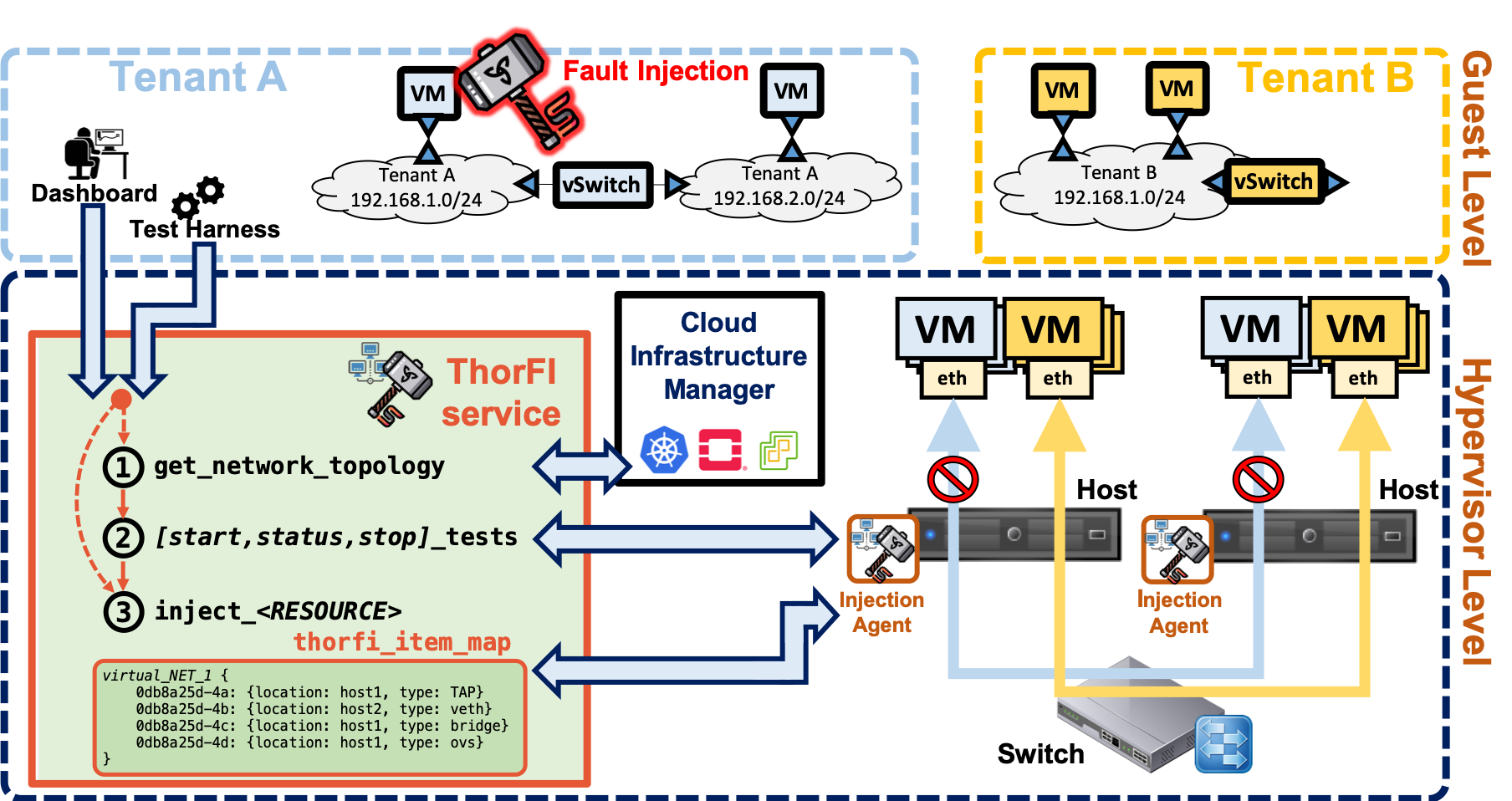}
  \caption{General architecture of the \toolname{} solution for network fault injection as-a-service.}
  \label{fig:netcast_overview}
\end{figure*}

\emph{Fault injection} is a significant solution to emulate these problems in a controlled way, to make distributed systems more fault-tolerant. For example, several large companies, such as Netflix, Uber, Amazon, have been using fault injection for their \emph{chaos engineering} and \emph{game day} exercises to assess the reliability of their services  \cite{allspaw2012fault,rosenthal2020chaos}. Unfortunately, fault injection has a high entry barrier, and it is still beyond the reach of the minor service providers, due to the cost and complexity of planning and orchestrate fault injection experiments. Research has been addressing this challenge by developing many solutions applicable both at compile- \cite{schwahn2018fastfi,wei2014quantifying} and run-time \cite{marinescu2009lfi,zhang2021maximizing}; domain-specific languages to tailor fault injection for the specific system \cite{hoara2005language,aliabadi2016fidl,cotroneo2020profipy,rodrigues2020model}; heuristics aim at minimizing the number of fault injection experiments to identify weak points in the system \cite{Gunawi2011a,Joshi2011b,alvaro2015lineage}; and automated techniques for analyzing fault injection data \cite{oliveira2011composed,cotroneo2020fault}. Despite these advances, it is still difficult to deploy fault injection toolchains into cloud infrastructures. 
Indeed, the existing solutions are difficult to use since they are restricted to run either at \emph{guest-level} (i.e., within the guest OS of VMs in the virtual network) or at the \emph{hypervisor-level}. 
To use guest-level injectors, the cloud tenants have to modify their VMs by installing tailored tools, which makes it more difficult to deploy injectors in large-scale systems. Moreover, it is not possible to install guest-level injectors if the VM runs images from commercial vendors (e.g., proprietary VNFs), or when there are security and privacy concerns. Instead, hypervisor-level injectors indiscriminately hit network traffic flows, regardless of which tenants are affected by the injection, since this kind of tool lacks high-level information about the tenant's network flows. 

In this work, we present a novel fault injection solution (\toolname{}) tailored for virtual networks deployed on cloud infrastructures. The proposed solution includes a unique architecture in the design space of fault injectors, as it meets both the objectives of (i) making fault injection cost-efficient to deploy for a cloud tenant, by avoiding the installation of guest-level injectors, and (ii) preventing injections from interfering with other tenants (e.g., injections in tenant A's virtual network do not impact on tenant B, \figurename{}~\ref{fig:netcast_overview}). The proposed architecture interacts with the cloud infrastructure management platform, in order to obtain high-level information about the virtual networks, such as virtual routers, and to map them to low-level resources, such as physical network interfaces, processes, and network flows. Then, \toolname{} orchestrates low-level injectors deployed across nodes in the data center, to ensure injecting faults only into the targeted virtual network, without affecting other tenants. 
Moreover, the proposed solution includes an implementation of the \toolname{} architecture integrated with OpenStack, a popular platform used in several private and public cloud computing infrastructures \cite{OpenStackUsers}, and the basis of over 30 commercial products \cite{OpenStackProducts}. We release this implementation as open-source software\footnote{\url{https://github.com/dessertlab/thorfi}}.

We evaluated the proposed solution in the context of two case studies (an NFV IP Multimedia Subsystem, and a cloud web application), by using fault injection for two relevant goals: (i) Extending functional test cases with faults; (ii) Evaluating quality of service under faults, in order to support capacity and redundancy planning. \revision{The experiments show that faults can indeed impact the functional behavior of the service (e.g., protocol compliance) in 4\%-34\% of the test cases, thus pointing out issues in the software implementation of the service. Moreover, the injection experiments on the cloud application highlight that the simple redundancy of network elements does not guarantee a high quality of service, since faults in one virtual network segment can propagate to other segments, and can degrade the throughput and response time of the cloud application as a whole, by about 3 times in the worst case. These results point out the need for careful capacity planning of redundancy, to be supported by solutions such as \toolname{}.}

The rest of the paper is organized as follows. Sections~\ref{sec:solution} and \ref{sec:implementation} present the design and implementation of \toolname{}. Section~\ref{sec:casestudy} presents the experimental results. Section~\ref{sec:discussion} provides discussion about qualitative evaluation of \toolname{}.
Section~\ref{sec:related} discusses related work on network fault injection. Section~\ref{sec:conclusion} concludes the paper.

\section{Related Work}
\label{sec:related}



The injection of faults to assess robustness and fault-tolerance under exceptional conditions is an established technique. We here review representative studies in the field of networked systems and virtualization technologies.

\noindent
$\rhd$ \textbf{Network fault injection.} 
Early network fault injection solutions were aimed at testing network protocol implementations.  \textit{VirtualWire} \cite{de2003virtualwire} injects network faults by triggering the injection when specific network events occur, as defined by the user via high-level specifications written in a declarative language. \textit{Farj et al.} \cite{farj2012fault} developed a fault injection tool for service-oriented web applications, which proxies SOAP requests and emulates faults occurring in wide-area networks, by corrupting SOAP messages. \textit{Olivera et al.} \cite{oliveira2010dependability} presented a tool that emulates partitions in IP-based networks, in order to evaluate the dependability of distributed applications. The solution includes a global coordinator, which orchestrates partitioning scenarios defined by the user; and communication fault injectors deployed across nodes, which use a Netfilter-based Linux kernel module to block network traffic. These early tools are not integrated with cloud infrastructures.

\noindent
$\rhd$ \textbf{Fault injection testing of cloud computing systems.} Fault injection has also been applied to test the robustness of software systems used in cloud computing infrastructures. 
\textit{Ju et al.} \cite{Ju2013a} tested the resilience of the OpenStack cloud computing platform, by injecting crashes (e.g., by killing VMs or service processes), network partitions (by disabling communication between two subnets), and network traffic latency and packet losses (by disrupting REST service requests). 
\emph{CloudVal} \cite{Pham2011} and \textit{Cerveira et al.} \cite{cerveira2015recovery, cerveira2020effects} use fault injection to test the isolation among hypervisors and VMs, by deliberately introducing CPU/memory corruptions and resource leaks. \textit{Pham et al.} \cite{pham2017failure} used fault injection against OpenStack, to understand the nature of its failures. They create signatures for each failure for supporting diagnosis when the same failures happen in production. However, these infrastructure-level solutions are not meant to be used by \emph{cloud tenants} for testing their own cloud applications.

\noindent
$\rhd$ \textbf{Fault injection in virtualized networks.} 
Fault injection has also been used to test applications deployed on virtualized networks. 
\emph{ChaosMonkey} \cite{chaos_monkey}, \emph{Jepsen} \cite{jepsen}, \revision{\emph{NEAT} \cite{alquraan2018analysis}, CoFI \cite{chen2020cofi},} and \textit{Gremlin} \cite{gremlin} are representative solutions for \textit{chaos engineering}, which test the resilience of distributed applications deployed on the cloud. These tools are deployed inside the same virtual machines that run the application (i.e., at guest-level), in order to inject faults such as network partitions, and network traffic losses, and latency. 
\emph{Fate} \cite{Gunawi2011a} and \emph{PreFail} \cite{Joshi2011b} simulate partitions and node crashes to test distributed recovery procedures, by injecting multiple faults during the same experiment. 
\textit{Chang et al.} \cite{chang2015chaos} proposed a fault injection tool for virtualized networks, based on SDN controllers. It injects faults by using the OpenVirteX hypervisor \cite{openvirtex} to re-configure the SDN controllers, such that they emulate network failure scenarios. Its objective is to verify that network invariants (e.g., reachability guarantees) still hold even in the presence of faults. 
In our previous work \cite{cotroneo2017nfv}, we proposed a benchmark to assess the dependability of NFV services. The benchmark guides how to systematically measure and evaluate dependability metrics, and comes with a suite of hypervisor-level fault injection tools, to emulate faults in the software, hardware, and network components.\\


\begin{table*}[h!]
\caption{\revision{Network fault injection solutions in the state-of-the-art compared to ThorFI}}
\label{tab:related_summary}
\resizebox{\textwidth}{!}{%
\sffamily 
\footnotesize 
\setstretch{0.90}
{\color{black}\begin{tabular}[h!]{C{2.5cm}C{3.5cm}C{3cm}C{5cm}C{1.4cm}C{3cm}C{2.5cm}}
\toprule\toprule
\normalsize\textbf{Study}  & 
\normalsize\textbf{Target} &
\normalsize\textbf{Fault Model} &

\normalsize\textbf{KPI} &

\normalsize\textbf{Multi-tenancy} &
\normalsize\textbf{Test automation} &

\normalsize\textbf{Intrusiveness} \\ \toprule\toprule


\textit{VirtualWire} \cite{de2003virtualwire} & 
    
    TCP and Rether \cite{chiueh1997fault} implementation in Linux 2.4.17 kernel & 
    \includegraphics[width=0.1\columnwidth]{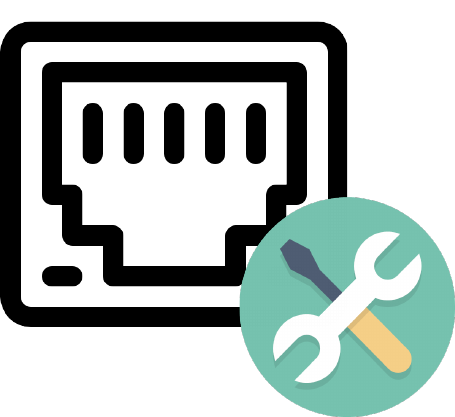}
    \includegraphics[width=0.1\columnwidth]{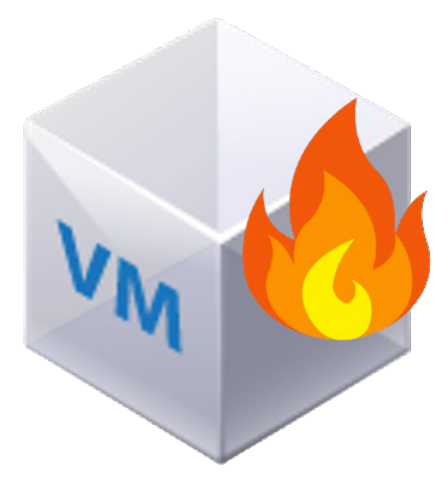} &
    \includegraphics[width=0.1\columnwidth]{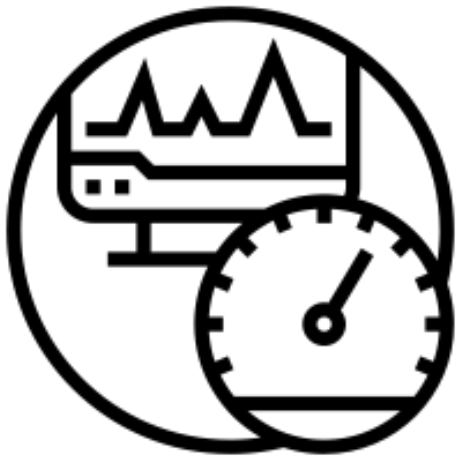}
    \includegraphics[width=0.1\columnwidth]{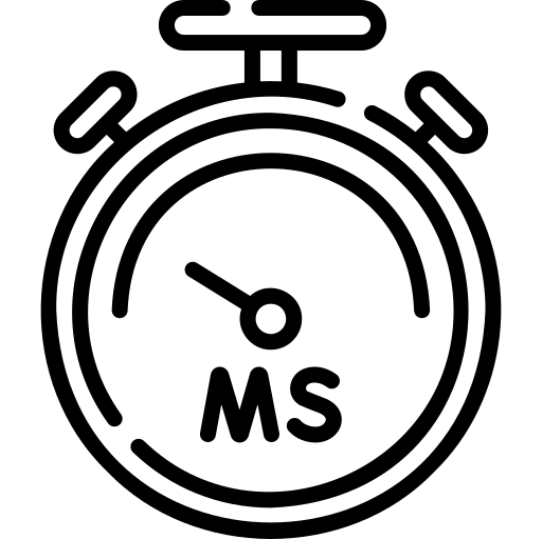} &
    \xmark &
    
    \includegraphics[width=0.1\columnwidth]{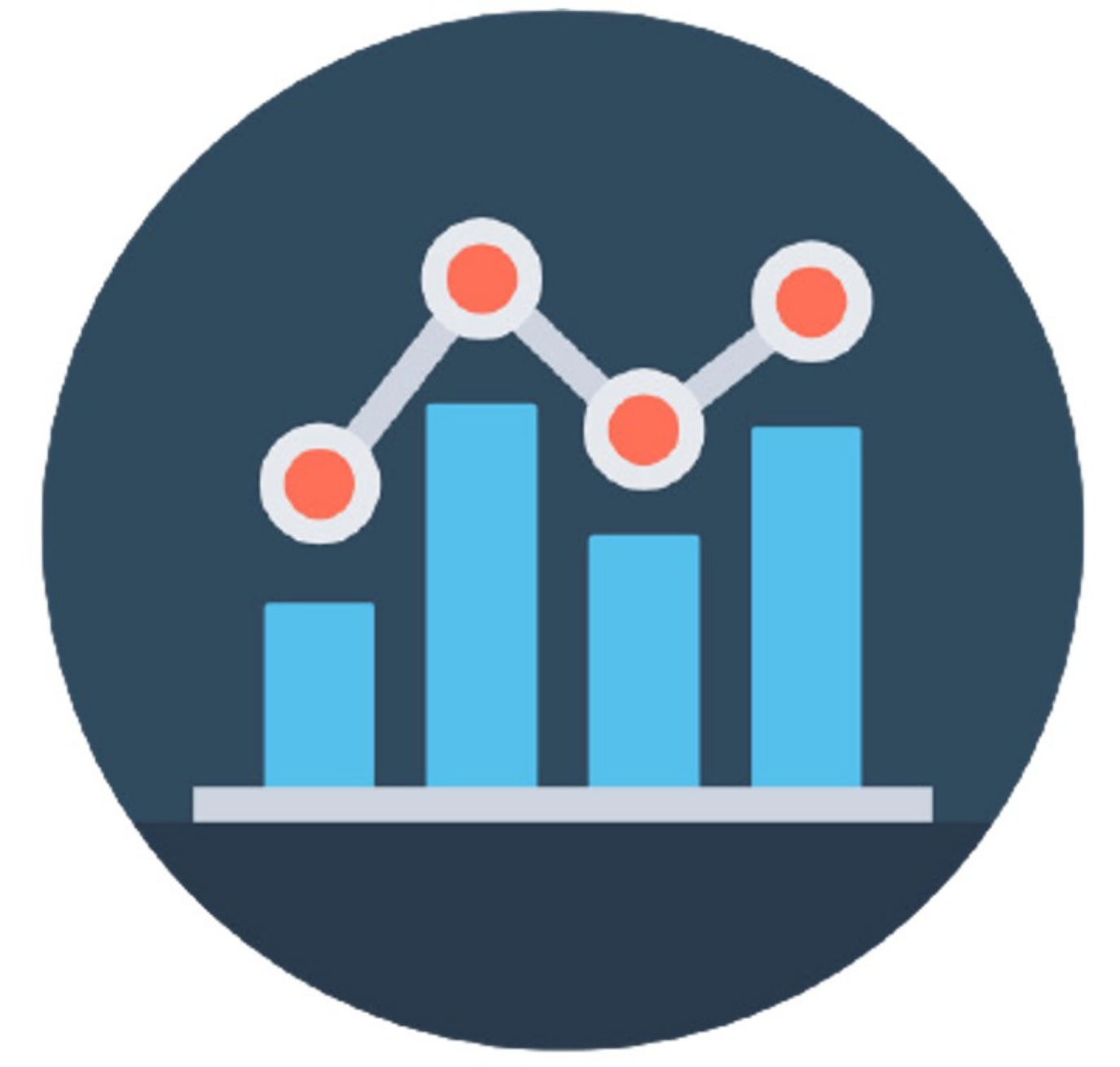} &
    
    \includegraphics[width=0.12\columnwidth]{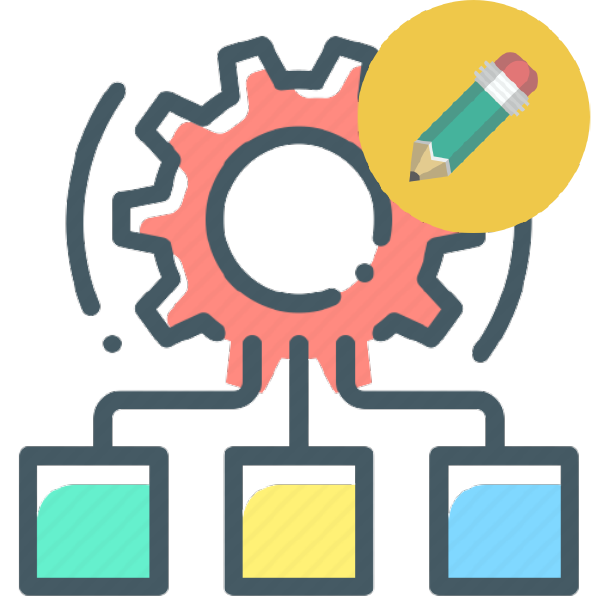} \\ \hline

\textit{Farj et al.} \cite{farj2012fault} & 
    
    Emulated network &  
    \includegraphics[width=0.1\columnwidth]{icon_fault_packet_flow_manipulation.pdf} &
    \includegraphics[width=0.1\columnwidth]{icon_kpi_latency.pdf} &
    \xmark &
    \xmark &
    \includegraphics[width=0.12\columnwidth]{icon_instrusiveness_hypervisor.pdf} \\ \hline

\textit{Olivera et al.} \cite{oliveira2010dependability} & 
    JGroups (toolkit for multicast communication) and PING & 
    \includegraphics[width=0.15\columnwidth]{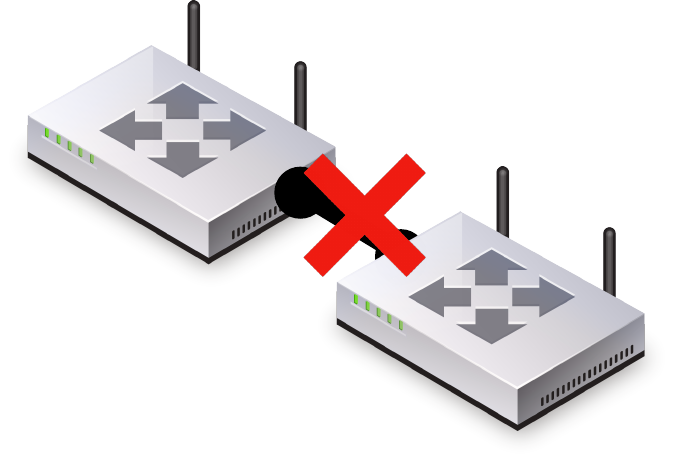} &
    \xmark &
    \xmark & 
    \includegraphics[width=0.1\columnwidth]{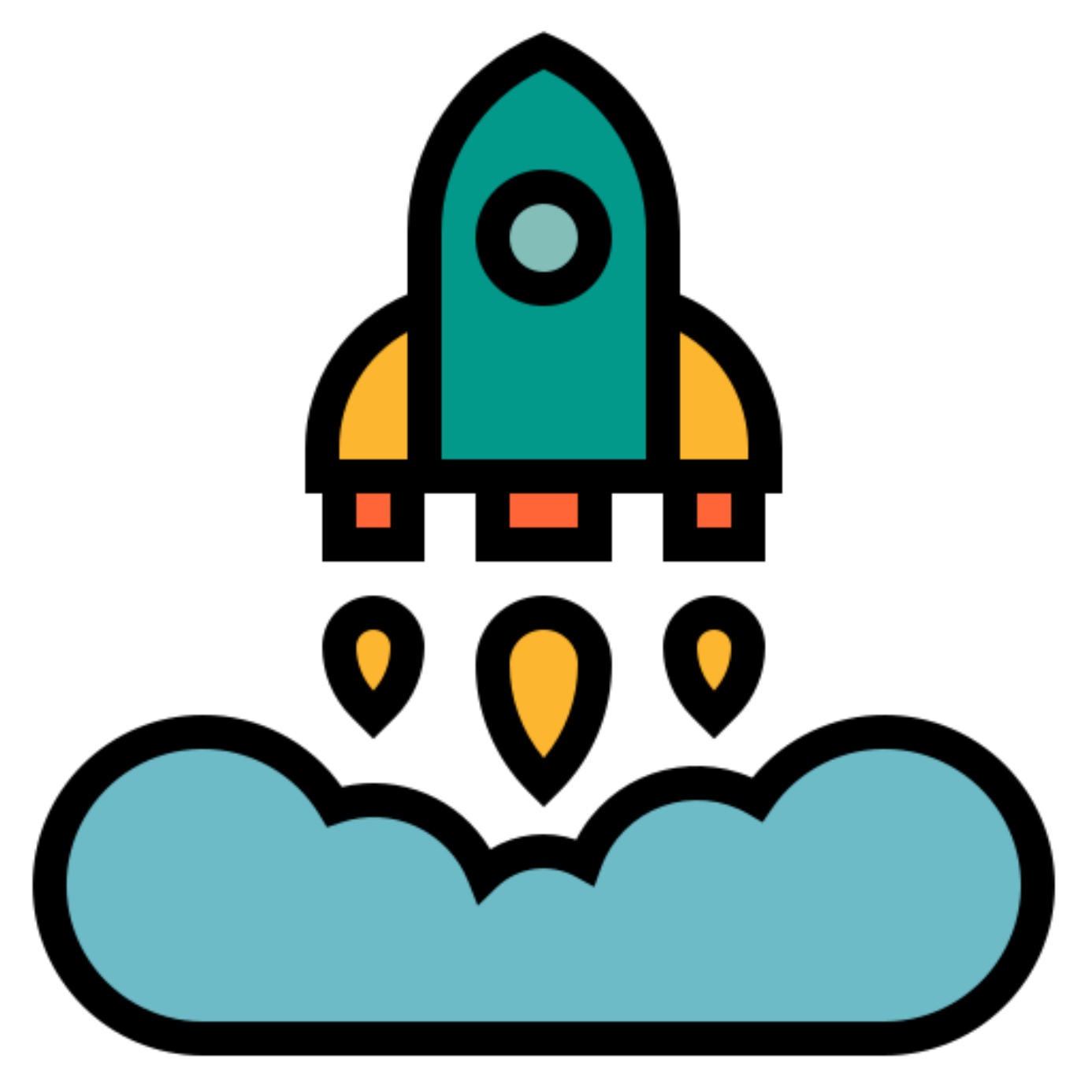} &
    
    
    \includegraphics[width=0.14\columnwidth]{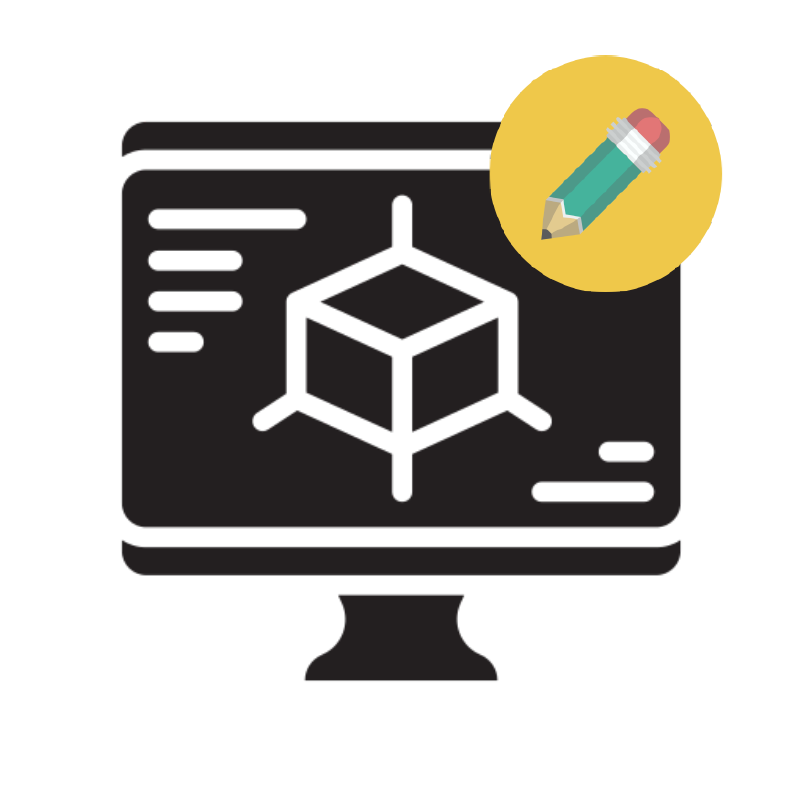} \\ \hline

\textit{Ju et al.} \cite{Ju2013a} & 
    OpenStack subsystems (Keystone, Glance, and Nova) &
    \includegraphics[width=0.15\columnwidth]{icon_fault_network_partition.pdf} &
    \includegraphics[width=0.1\columnwidth]{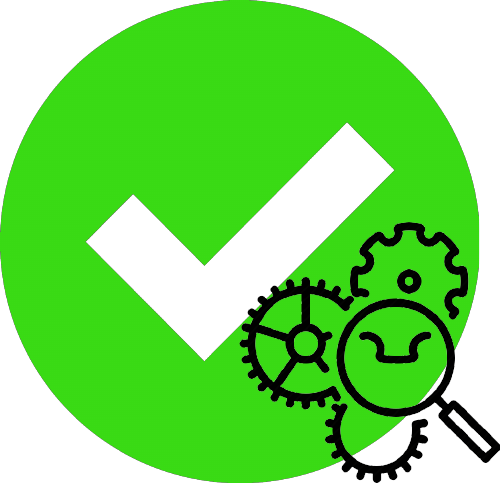} &
    \cmark &
    \includegraphics[width=0.08\columnwidth]{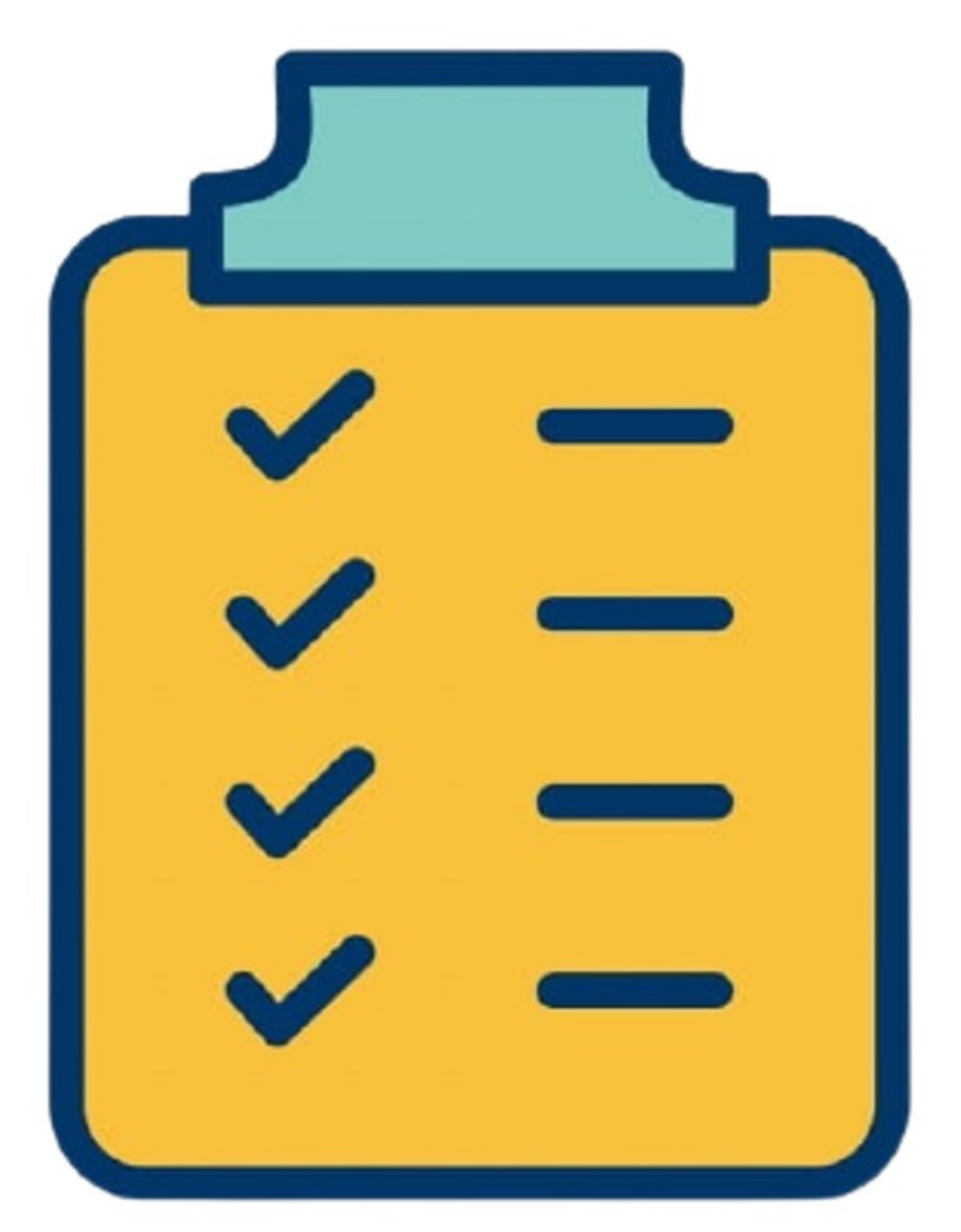}
    \includegraphics[width=0.1\columnwidth]{icon_test_deployment.pdf} &
    
    \includegraphics[width=0.14\columnwidth]{icon_instrusiveness_guest_mod.pdf} \\ \hline

\emph{ChaosMonkey} \cite{chaos_monkey} & 
    
    Spinnaker-based cloud deployments \cite{chaos_monkey_spinnaker} &
    
    \includegraphics[width=0.1\columnwidth]{icon_fault_packet_node_crash.pdf}
    \includegraphics[width=0.1\columnwidth]{icon_fault_packet_flow_manipulation.pdf}
    &
    
    \includegraphics[width=0.1\columnwidth]{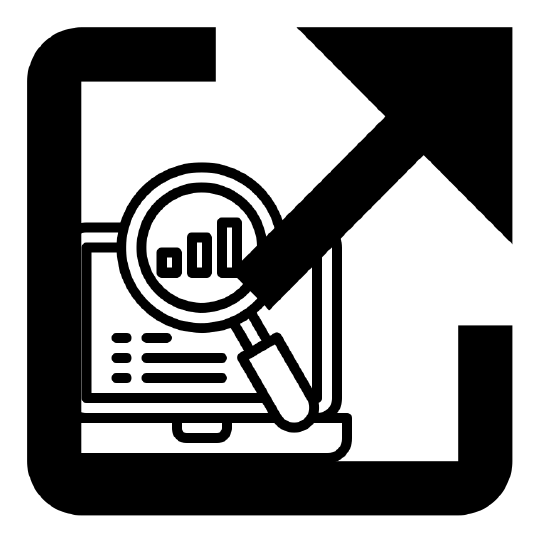} &
    \cmark &
    \includegraphics[width=0.08\columnwidth]{icon_test_planning.pdf}
    \includegraphics[width=0.1\columnwidth]{icon_test_deployment.pdf} &
    
    \includegraphics[width=0.14\columnwidth]{icon_instrusiveness_guest_mod.pdf} \\ \hline

\emph{Jepsen} \cite{jepsen} & 

    Several databases, coordination services, and queues management systems \cite{jepsen_analysis} &
    
    \includegraphics[width=0.15\columnwidth]{icon_fault_network_partition.pdf} 
    \includegraphics[width=0.1\columnwidth]{icon_fault_packet_flow_manipulation.pdf} &
    \includegraphics[width=0.1\columnwidth]{icon_kpi_func_correctness.pdf} &
    \xmark &
    \includegraphics[width=0.1\columnwidth]{icon_test_deployment.pdf} &
    
    
    \includegraphics[width=0.14\columnwidth]{icon_instrusiveness_guest_mod.pdf} \\ \hline


\emph{NEAT} \cite{alquraan2018analysis} & 
    Several distributed systems \cite{alquraan2018analysis} &
    \includegraphics[width=0.15\columnwidth]{icon_fault_network_partition.pdf} &
    \includegraphics[width=0.1\columnwidth]{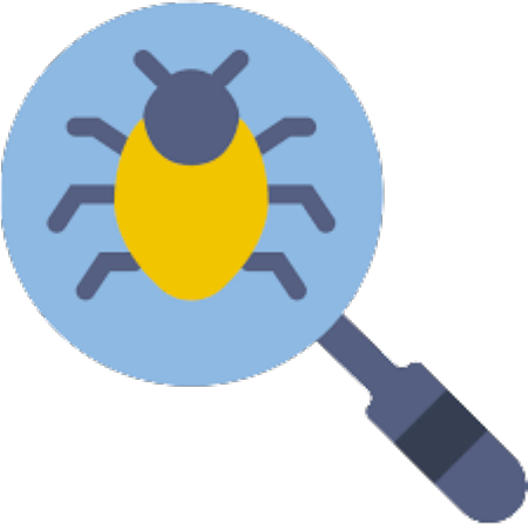} &
    \xmark &
    \includegraphics[width=0.1\columnwidth]{icon_test_deployment.pdf} &  
    \includegraphics[width=0.12\columnwidth]{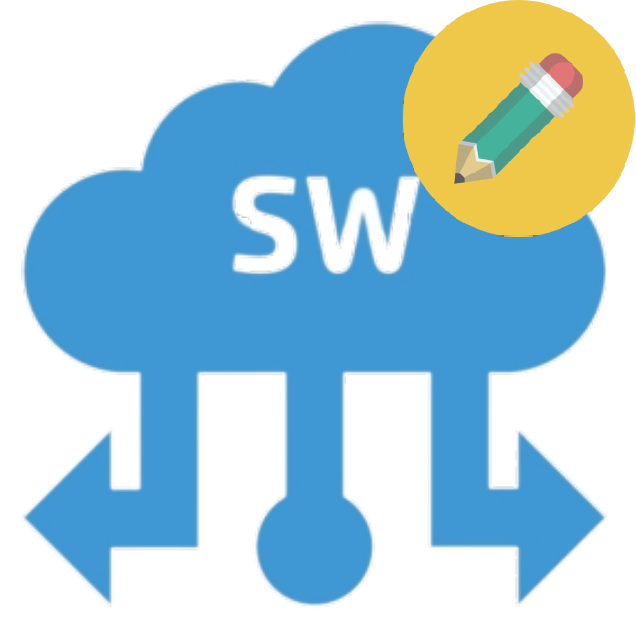}
    \includegraphics[width=0.14\columnwidth]{icon_instrusiveness_guest_mod.pdf} \\ \hline
    
\emph{CoFI} \cite{chen2020cofi} & 
    Cassandra \cite{lakshman2010cassandra}, HDFS \cite{shvachko2010hadoop}, and YARN \cite{apache_yarn} &
    \includegraphics[width=0.15\columnwidth]{icon_fault_network_partition.pdf} &
    \includegraphics[width=0.1\columnwidth]{icon_kpi_dependability.pdf} &
    \xmark &
    \includegraphics[width=0.08\columnwidth]{icon_test_planning.pdf}
    \includegraphics[width=0.1\columnwidth]{icon_test_deployment.pdf} &
    
    
    
    \includegraphics[width=0.14\columnwidth]{icon_instrusiveness_guest_mod.pdf} \\ \hline


\textit{Gremlin} \cite{gremlin} & 
    

    Several customers use Gremlin in their systems \cite{gremlin_customers}. Gremlin targets either hosts or containers. &
    \includegraphics[width=0.1\columnwidth]{icon_fault_packet_flow_manipulation.pdf} &
    \includegraphics[width=0.1\columnwidth]{icon_kpi_external_mon.pdf} &
    \cmark &
    \includegraphics[width=0.1\columnwidth]{icon_test_deployment.pdf} &

    
    \includegraphics[width=0.14\columnwidth]{icon_instrusiveness_guest_mod.pdf} \\ \hline

\emph{Fate} \cite{Gunawi2011a} and \emph{PreFail} \cite{Joshi2011b} & 
    
    HDFS \cite{shvachko2010hadoop}, Zookeeper \cite{hunt2010zookeeper}, Cassandra \cite{lakshman2010cassandra} &
    \includegraphics[width=0.15\columnwidth]{icon_fault_network_partition.pdf} &
    \includegraphics[width=0.1\columnwidth]{icon_kpi_dependability.pdf} &
    
    \xmark &
    
    \includegraphics[width=0.1\columnwidth]{icon_test_deployment.pdf} &
    
    \includegraphics[width=0.14\columnwidth]{icon_instrusiveness_guest_mod.pdf} \\ \hline

\textit{Chang et al.} \cite{chang2015chaos} & 
    SDN testbed &
    \includegraphics[width=0.1\columnwidth]{icon_fault_packet_flow_manipulation.pdf} &
    \xmark &
    \xmark &
    \xmark & 
    

    \includegraphics[width=0.12\columnwidth]{icon_instrusiveness_sdn_apis.pdf} \\ \midrule \midrule
    
\textbf{ThorFI} & 
    OpenStack-based cloud applications &
    
    \includegraphics[width=0.15\columnwidth]{icon_fault_network_partition.pdf}
    \includegraphics[width=0.1\columnwidth]{icon_fault_packet_flow_manipulation.pdf}
    \includegraphics[width=0.1\columnwidth]{icon_fault_packet_node_crash.pdf} &
    \includegraphics[width=0.1\columnwidth]{icon_kpi_throughput.pdf}
    \includegraphics[width=0.1\columnwidth]{icon_kpi_latency.pdf}
    \includegraphics[width=0.1\columnwidth]{icon_kpi_func_correctness.pdf}
    \includegraphics[width=0.1\columnwidth]{icon_kpi_dependability.pdf} &
    
    \cmark &
    
    \includegraphics[width=0.08\columnwidth]{icon_test_planning.pdf}
    \includegraphics[width=0.1\columnwidth]{icon_test_deployment.pdf}
    \includegraphics[width=0.1\columnwidth]{icon_test_reporting.pdf} &
    
    \includegraphics[width=0.12\columnwidth]{icon_instrusiveness_hypervisor.pdf} \\ \hline \hline
    
\end{tabular}}
}%

\includegraphics[width=2\columnwidth]{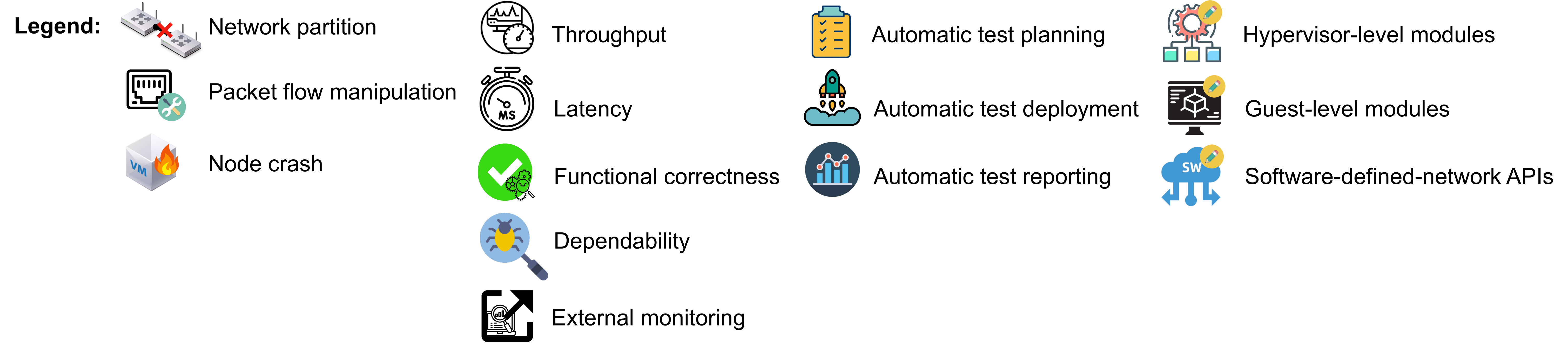}

\end{table*}


\revision{\tableautorefname{}~\ref{tab:related_summary} shows a comparison between \toolname{} with respect to the state of the art. We compared the studies focusing on key aspects for fault injection, including the target, the adopted fault models, the key performance indicators used for the provided experimentation (e.g., throughput, latency, availability, functional correctness), the support for multi-tenancy scenarios, support for automation beyond the simple injection of faults (e.g., support for deploying the system under test, and for managing and reporting large sets of tests), the intrusiveness of the solutions (e.g., the need to install special software at guest-level, hypervisor-level, or software-defined-network controller). We want to remark that the aim of this work is not to introduce new network fault models as they are well established in the fault injection literature. However, as the reader can notice, the proposed approach can cover all the existing network fault models. Further, it advances the state-of-the-art by providing a way to automatically manage and orchestrate fault injection experiments against complex virtualized network topologies underlying running applications in multi-tenancy scenarios without modification at the guest level. Indeed, ThoFI requires installing modules only at the hypervisor level, which include the front-end agent (running on a controller node in the cloud infrastructure) and the injection agents (running in each compute/network node belonging to the targeted tenant). Further,} differing from traditional hypervisor-level solutions, our one assures tenant isolation, by selectively injecting faults only on the virtual resources of a specific tenant. Moreover, \toolname{} has been designed to be integrated with the typical use cases for fault injection on the cloud, i.e., extending functional testing, and performing availability and performance measurements for cloud applications. \revision{Finally, ThorFI is released as open-source software publicly available.}


\section{\toolname{} Design}
\label{sec:solution}

In this section, we present the high-level architecture of \toolname{} (\S{}~\ref{subsec:thorfi_architecture}), how it integrates with the main use cases for fault injection on the cloud (\S~\ref{subsec:use_cases}), and which network faults it can inject (\S~\ref{subsec:faults}).

\subsection{Architecture}
\label{subsec:thorfi_architecture}

\toolname{} is an ``as-a-service'' solution, which allows tenants to perform fault injection tests on their virtual networks. \toolname{} has been designed to perform fault injection without modifying the tenants' virtual machines (e.g., for security and privacy reasons), by running at the hypervisor level. At the same time, \toolname{} is still able to selectively inject faults only for a specific tenant, while other tenants are \emph{isolated} from the effects of the injection.

\figurename{}~\ref{fig:netcast_overview} shows the architecture of \toolname{}, in the context of two tenants on the same cloud infrastructure. A tenant (e.g., \emph{Tenant A} in the figure) generates a fault scenario to test its cloud application. For example, the tenant wants to simulate a network partition between virtual storage nodes, e.g., to test the impact of data inconsistency between replicas. The tenant does not need to install any fault injection software in its virtual network. Instead, it suffices to request a fault scenario to the \toolname{} \textit{service} (see \figurename{}~\ref{fig:netcast_overview}). 
The service automatically maps the fault scenario (at the guest level) to fault injection actions to be performed (at the hypervisor level). The service deploys fault injection agents across the physical hosts of the infrastructure, and selectively inject faults only in the traffic of \emph{Tenant A}. Instead, the network traffic of \emph{Tenant B} is not affected by fault injection. The tenant does not need to know where its virtual resources are deployed on the physical infrastructure.

The \toolname{} service provides a set of 55 REST APIs for managing and orchestrating fault injection experiments. 
\toolname{} gives clients choice to orchestrate their experiments either by fully using provided REST APIs (step \circled{1}, \circled{2}, and \circled{3} in \figureautorefname{}~\ref{fig:netcast_overview}) or by exploiting their own orchestration software and invoking only \toolname{} injection REST APIs (step \circled{3} in \figureautorefname{}~\ref{fig:netcast_overview}). In particular, \toolname{} provides the {\lmttfont get\_network\_topology} REST API (step \circled{1} in \figureautorefname{}~\ref{fig:netcast_overview}) to retrieve information about virtual network resources that are potential targets for injections, including their topology and IDs.
Then, the client (as discussed in the next subsection) can define a \textit{fault injection test plan}, by selecting which virtual network resources should be injected. The test plan lists a set of tests and, for each test, \emph{what} fault should be injected, \emph{where} in the network it should be injected, and \emph{when} it should be injected during the test. The client then uses the {\lmttfont start\_tests}, {\lmttfont status\_tests}, and {\lmttfont stop\_tests} REST APIs (step \circled{2} in \figureautorefname{}~\ref{fig:netcast_overview}), which respectively run, provide current status, and stop the execution of a fault injection test plan. 
The {\lmttfont start\_tests} REST API uses, in turn, the {\lmttfont inject\_\textit{<RESOURCE>}} REST APIs (discussed in the following) to actually inject the faults. After each test has been executed, \toolname{} invokes an internal method, {\lmttfont save\_logs}, to collect data about service performance and availability, later elaborated to provide concise reports to the clients. 
These REST APIs hold various information (e.g., the status of the tests) in a private database of \toolname{}.

About the injection functionalities, as mentioned before, \toolname{} provides a set of REST APIs named {\lmttfont inject\_\textit{<RESOURCE>}} (step \circled{3} in \figureautorefname{}~\ref{fig:netcast_overview}), where {\lmttfont <RESOURCE>} represents any of the virtual network resources, such as virtual switches and virtual routers. Each virtual network resource is \emph{mapped} to a set of network interfaces at the hypervisor level. Thus, \toolname{} abstracts the virtual network resources as a set of {\lmttfont thorfi\_item}s, i.e., the network interfaces to be actually targeted by fault injection. Each {\lmttfont thorfi\_item} is characterized by \textit{i)} an \textit{ID}, \textit{ii)} a \textit{location} (i.e., the physical machine that hosts the network interface), and \textit{iii)} a \textit{type} (e.g., tap devices, veth pairs, Linux bridges, Open vSwitch bridges, and so on). \toolname{} stores a hash table, namely {\lmttfont thorfi\_item\_map}, that maps each virtual network resource with its underlying {\lmttfont thorfi\_item}s. Internally, \toolname{} invokes the {\lmttfont{do\_injection\_thorfi\_item}} method, which uses the {\lmttfont thorfi\_item\_map} to identify the target network interfaces and to start injection by communicating with \toolname{} \emph{\injectorAgent{}s}. 
We remark that \toolname{} is flexible since it allows clients to get network topology and orchestrate configured test plans also with their customized software. Indeed, clients can bypass {\lmttfont get\_network\_topology} and {\lmttfont [start, status, stop]\_tests} REST APIs, and directly invoke {\lmttfont inject\_\textit{<RESOURCE>}} REST APIs on recognized target virtual resources.

\begin{figure}[!h]
  \centering
  \includegraphics[width=\columnwidth]{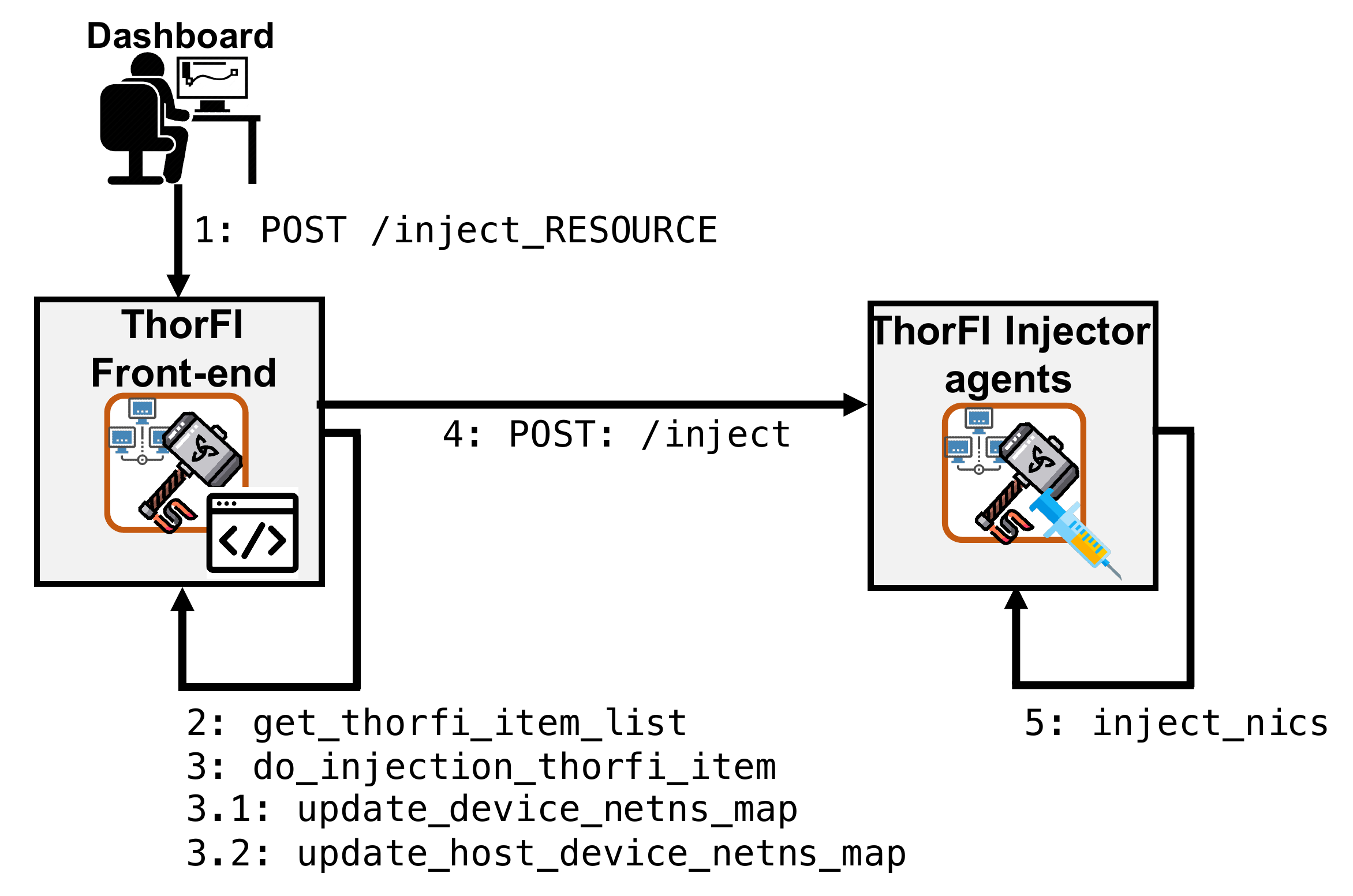}
  \caption{\revision{Example of ThorFI sequence workflow for injecting resource faults.}}
  \label{fig:thorfi_sequence_workflow}
\end{figure}

The REST APIs described above are implemented in the \toolname{} service, according to how the target cloud-computing platforms manage networking resources. The \toolname{} service interacts with \textit{injection agents} running in the machines across the data center to actually inject network faults. \revision{\figureautorefname{}~\ref{fig:thorfi_sequence_workflow} shows an example of the sequence of events in \toolname{} after invoking the {\lmttfont inject\_\textit{<RESOURCE>}} REST API. More details are described in the documentation of \toolname{}.} 

\begin{figure*}[!h]
  \centering
  \includegraphics[width=2\columnwidth]{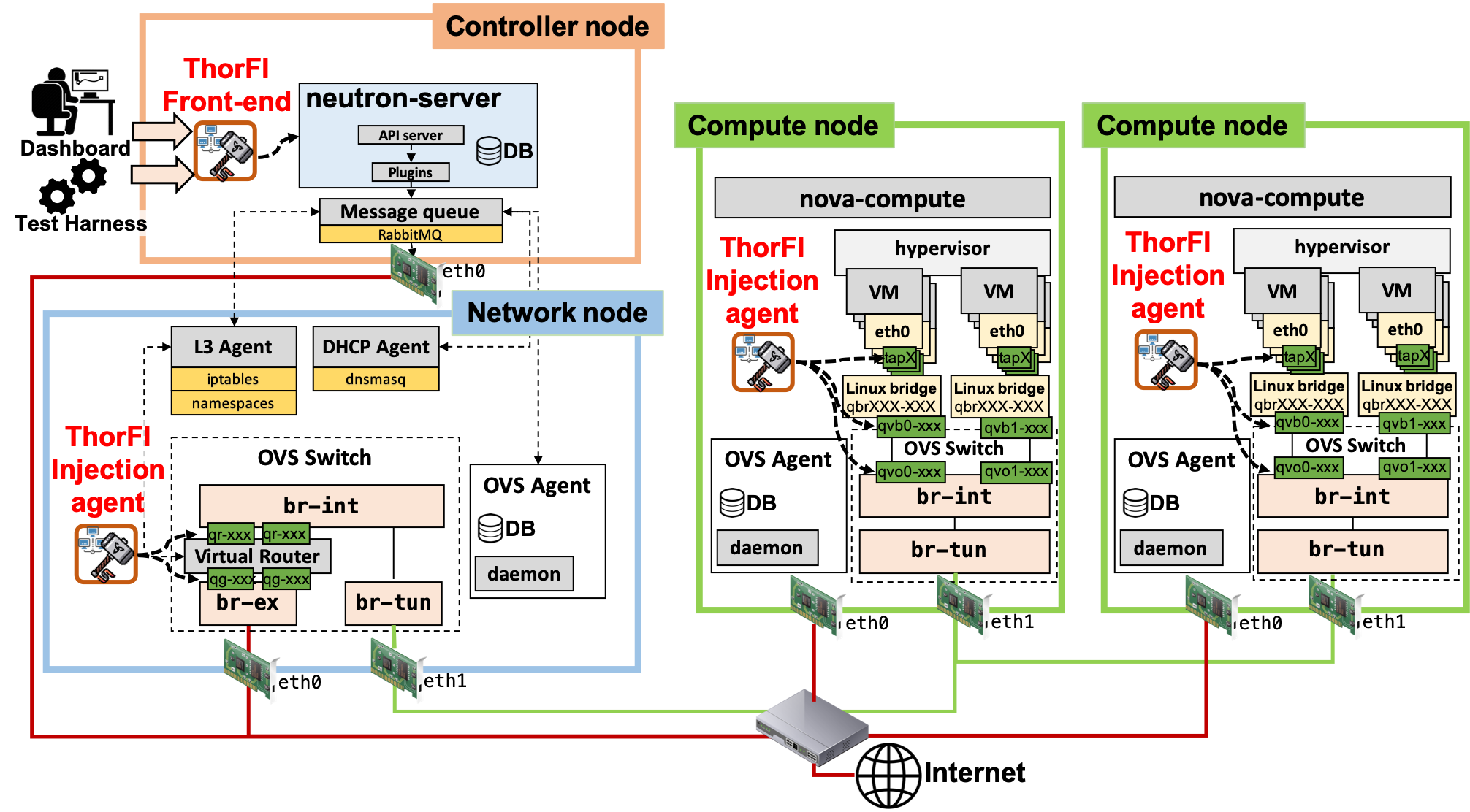}
  \caption{The proposed solution in the context of OpenStack.}
  \label{fig:openstack_networking}
\end{figure*}

The proposed architecture in \figureautorefname{}~\ref{fig:netcast_overview} is agnostic to cloud infrastructure managers and hypervisors. Without loss of generality, in this work, we present an implementation of \toolname{} for the OpenStack cloud computing platform. A discussion on the portability of \toolname{} is provided in \sectionautorefname{}~\ref{sec:discussion}. \figurename{}~\ref{fig:openstack_networking} shows the proposed solution in the context of OpenStack.

The user interacts with \toolname{} either through a web dashboard interface or through a command-line interface. In both cases, the interface communicates with a \textit{\netcastAgent{}} provided by the \toolname{} service (the \toolname{} \textit{Front-end} in \figurename{}~\ref{fig:openstack_networking}), which orchestrates the fault injection experiments. In turn, the \textit{\netcastAgent{}} service interacts with \toolname{} \emph{\injectorAgent{}s}, which are deployed on the physical machines across the data center, and perform the actual injection of faults. Moreover, the \textit{\netcastAgent{}} interacts with the cloud infrastructure manager, in order to collect information about the tenant and its virtual networks. The \textit{\netcastAgent{}} uses this information to inject faults in selected parts of the infrastructure, such that the injections only affect the targeted virtual network, without impacting on other tenants.

\subsection{Use Cases}
\label{subsec:use_cases}

\toolname{} is designed to support two main use cases of fault injection for cloud systems: (i) to provide end-to-end, self-contained automation of performance and availability evaluation experiments, and (ii) to integrate fault injection with existing functional test suites.

In the first use case, \toolname{} is adopted for the \textbf{redundancy and capacity planning} of applications deployed on the cloud. In order to achieve high availability and performance, such applications are often deployed by using replicated virtual machines and subnets, and by dividing the workload through balancers. It is tricky for system designers to plan redundancy and capacity to meet both cost and performance goals at the same time. Benchmarking tools allow system designers to assess which performance levels (e.g., throughput) can be achieved with a given amount of resources (i.e., cost). However, these tools alone cannot assess whether performance levels still hold under faults, which can impact severely due to the lower amount of available resources. Our proposed solution can execute performance benchmarking experiments enriched with injected faults.

In the second use case, \toolname{} is adopted for \textbf{extending functional tests with fault injection}. In functional testing, a client (\emph{test harness}) exercises the system under test (SUT) by submitting a set of diverse inputs, in order to assess whether the system correctly implements its specification. For example, in the context of NFV, the softwarized network functions have to comply with industry standards, such as the SIP protocol in the case of the IP Multimedia Subsystem \cite{rosenberg2002sip}. However, specifications entail many corner cases, which often can not be reproduced with the sole inputs from the test harness. Fault injection is useful to make functional testing more thorough, by introducing anomalous events to cover more corner cases. Our fault injection solution can be integrated with a test harness.

\noindent
$\rhd$ \textbf{End-to-end performance and availability evaluation.} 
In this first use case, the user interacts with \toolname{} to configure all the steps to automate fault injection tests. These steps are repeated for every fault listed in a \textit{fault injection test plan}. The steps include:

\begin{enumerate}
    \item to automatically (re-)deploy workload generators (e.g., simulated clients) on the virtual network;
    \item to execute the workload generators to stimulate the SUT;
    \item to inject one fault from the test plan; 
    \item to synchronize with the completion of the workload and of the SUT;
    \item to collect data from the workload generators and the SUT, including measurements of service performance and service errors reported by workload generators.
\end{enumerate}

To configure this workflow, we expect the user to interact with \toolname{} through a web dashboard. Despite \toolname{} provides also a command-line interface, the dashboard is important to make fault injection a cost-efficient process, by abstracting virtual networks from the corresponding physical resources where faults are injected. It visualizes the virtual network as a graph, where nodes represent network elements (e.g., network interfaces, virtual routers, subnets, etc.). The user can interact with the graph, by selecting a node, and by using a wizard to select which fault to inject in that network element. In this way, the user can annotate the graph element with one or several faults, in order to create a fault injection test plan. The injection of faults is further discussed in \S{}~\ref{subsec:faults}.

Then, the user configures the deployment of workload generators on the existing virtual network. The user can either configure its own workload generator, by providing the image of a pre-configured virtual machine; or, it can reuse images with pre-configured workload generators provided by \toolname{}. The dashboard visualizes again a graph representing the virtual network. This time, the user interacts with the graph to add new nodes (representing instances of the workload generator), and to attach them to the virtual network. For example, the user can add an HTTP workload generator to the side of the network exposed to the public internet, to simulate web clients. As another example, the user can use a workload that runs on two nodes (e.g., a client and a server) at different sides of the virtual network, to assess network performance between the two sides. 

Finally, \toolname{} collects data from the workload generators for reporting purposes. Then, it computes performance and availability metrics, such as throughput, latency, error rate as perceived by the workload generators. These metrics are initially reported in aggregate form, by plotting the average and standard deviation for each experiment. To provide context for these results, \toolname{} also reports the values of these metrics measured under fault-free (no fault injected) executions of the workload. Thus, the user can identify which fault injection tests caused drops in performance and availability compared to fault-free executions. Moreover, \toolname{} allows the user to drill down the data, by plotting detailed time series with performance and availability measures throughout the experiment. The tool annotates the time series with visual clues to show when a fault has been injected during the experiment. The tool also allows the user to download the raw data from the workload generators for further analysis, e.g., to inspect error messages reported by the client.

\noindent
$\rhd$ \textbf{Extending existing functional testing suites with fault injection.} 
\toolname{} also supports a second type of workflow, which integrates fault injection with existing testing suites. In this scenario, the user already has a test automation toolchain in place, which automatically deploys the SUT and exercises it with functional and regression tests. For example, a testing workflow for NFV software deploys one or more Virtual Network Functions (VNF) to form a service chain, and generates input traffic to test the network. 

The user can extend existing functional test cases with fault injection. In this case, faults are injected in order to trigger fault-tolerance mechanisms during the execution of the tests. The user can assess whether the functional behavior of the SUT is still correct even in the case of faults. The user can extend the test cases by invoking \toolname{} from the testing toolchain, either by using the command-line interface or by using directly the injection REST APIs. The testing toolchain should first call \toolname{} to schedule the injection of a fault with a timer; then, it can perform the test case, and the fault will be injected during the execution of the test. Another approach is to call \toolname{} at a given point of the test case, in order to trigger a fault at that point of the test. In both cases, the test suites can be extended with fault injection with minimal changes to the existing test harness.

\subsection{Injected Faults}
\label{subsec:faults}

\toolname{} supports the injection of several types of network faults. To inject faults into the virtual network of a specific tenant, the \netcastAgent{} retrieves information about which hypervisor-level resource serves the tenant: for example, which network nodes route the virtual traffic; and which network interfaces and processes are allocated to the virtual network. The injections are actually performed by the injection agents installed on the nodes of the data center. The agents inject faults by changing the configuration of the infrastructure at the hypervisor level: for example, by deallocating a resource (e.g., to emulate the outage of a virtual router or subnet), or by setting the loss or corruption rate of a network interface.

\toolname{} allows the user to configure fault injection with respect to the following aspects:

\begin{itemize}
\item \textbf{Fault type}: Two main categories of faults can be injected. The first category is represented by faults at the network traffic level, which includes the loss, delay, corruption, duplication of network packets, as well as artificial network limit rate \cite{cotroneo2017nfv}. The second category is the injection of faults on the virtual configuration, by forcing the unavailability of virtual network resources (e.g., virtual routers). These faults selectively remove elements from the virtual network (along with other elements that depend on them) and automatically restore them after fault injection. These faults can be caused by buggy updates, mistaken re-configurations, or other sporadic actions by system administrators \cite{fu2017runtime,lu2013cloud}, and network partition issues \cite{oliveira2010dependability}.

\item \textbf{Fault intensity}: Depending on the fault type, the user can tune the extent of traffic that is injected (in terms of percentage), and the magnitude of the fault effects, such as, the amount of delay in milliseconds for latency injections, and the amount of modified data for corruption faults. This applies only to network traffic level faults.

\item \textbf{Fault pattern}: The timing pattern of the injection. It can be \emph{random} (faults are injected in random packets); \emph{persistent} (faults are continuously injected during a period); \emph{bursty} (faults are periodically injected); \emph{degradation} (the amount of injected packets increases over time). This applies only for network traffic level faults.

\item \textbf{Fault target}: The protocol that should be injected. The user can select the target among a set of standard protocols, both at the network level, such as HTTP, and at the application level, such as database and key-value store traffic. The user can also configure non-standard protocols. This applies only to network traffic level faults.

\item \textbf{Fault injection timing}: Fault injection timings include the time to wait before injecting a fault (\textit{pre-injection time}), the time for which the fault remains activated (\textit{injection time}), and the time to wait before completing the fault injection experiment after the injection time (\textit{post-injection time}). Pre- and post-injection times are commonly used to warm up the SUT and wait for (potential) recovery after errors/failures respectively.

\end{itemize}

\section{OpenStack Implementation}
\label{sec:implementation}


The design of \toolname{} applies to several cloud computing platforms and virtualization technologies. 
We implemented \toolname{} in the context of the OpenStack cloud management platform, since it is widespread among public cloud infrastructure providers, private users, and commercial products \cite{OpenStackUsers,OpenStackProducts}. \toolname{} is implemented in Python and consists of $\sim5.5$ KLOCs. \revision{Our implementation integrates popular low-level fault injection tools for Linux/KVM- and Xen-based servers (e.g., {\lmttfont tc} tool \cite{tc_tool}, and can be ported to other kinds of virtualization approaches like container-based (e.g., Docker and Kubernetes). The design of \toolname{} is generalized enough to apply also to other cloud management platforms and hypervisors, such as VMware vSphere and ESXi, by integrating other low-level fault injection tools, such as those from previous studies \cite{cotroneo2017nfv,chaos_monkey,Pham2011,cerveira2015recovery}.} In the following, we provide more details about the implementation and how to integrate the design to the OpenStack cloud management platform.





The components of \toolname{} (the \netcastAgent{}, and one or more instances of the \injectorAgent{}) are meant to be deployed across the servers of the data center managed by OpenStack. OpenStack deployments organize servers into clusters, where each cluster includes three types of nodes \cite{openstack_deployments}: \emph{(i)} a \textit{Controller} node, where most of the shared OpenStack services and other tools run. The Controller node supplies API, scheduling, the image store, the identity service, and other shared services for the cloud. The \emph{Nova} compute management service and the \emph{Neutron} server are also configured in this node; \emph{(ii)} a \emph{Network} node, which provides virtual networking and networking services to Nova instances using the Neutron Layer 3 and DHCP network services; and \emph{(iii)} one or more \textit{Compute} nodes, where Nova compute instances are installed using the hypervisor component of the \textit{OpenStack Compute} service. 
The \toolname{} \netcastAgent{} runs in the controller node, in order to retrieve information about the virtual network configuration and instances. The \toolname{} \injectorAgent{}s run on each network and compute node in the deployment, in order to perform fault injection actions there.

\subsection{Injection of Network Traffic Faults}
\label{subsec:network_traffic_faults}

\toolname{} uses OpenStack REST APIs to discover the virtual network resources that will be targets for injections. The \toolname{} \netcastAgent{} uses the {\lmttfont list} REST APIs \cite{openstack_networking_API} to build the virtual network topology, where users can configure injections for each network element. \toolname{} \netcastAgent{} provides an injection API for each virtual network resource type, including virtual networks, subnets, routers, floating IPs, and ports (e.g., {\lmttfont /inject\_subnet} APIs). 

Each network element is associated with one or more {\lmttfont thorfi\_item}s within the {\lmttfont thorfi\_item\_map} (see \S~\ref{subsec:thorfi_architecture}). Specifically, {\lmttfont thorfi\_item}s in OpenStack are actually \emph{network ports} on a specific physical host in the datacenter. The network port is the basic unit for OpenStack networking and represents a connection point for attaching a device, such as a NIC of a server, the interfaces of a router, and so on. Given a target network resource (i.e., a virtual router), \toolname{} iterates over the network ports that refer to the target virtual resource to perform fault injection.

To inject faults at network traffic level (packet drop, delay, and corruption), \toolname{} uses {\lmttfont tc} as a low-level fault injection tool \cite{tc_tool}. \toolname{} differentiates injections based on the port types (the {\lmttfont device\_owner} field in OpenStack).


\noindent
$\rhd$ \textbf{Virtual NICs}. The traffic over a virtual NIC of a specific VM instance can be injected through network ports identified as {\lmttfont compute:nova} type. This port represents a \textit{tap interface} on the Compute node running the instance \cite{tap_interface}. In turn, the tap interface is attached to the \emph{integration} Linux bridge ({\lmttfont br-int}), which performs VLAN tagging, and to the \emph{tunnelling} bridge ({\lmttfont br-tun}), which connects the Compute node to the Network node through GRE tunnels (right side of \figurename{}~\ref{fig:openstack_networking}). To inject faults, the \toolname{} \netcastAgent{} matches the port ID in OpenStack of the target virtual NIC with the underlying tap device ID. Then, it sends an injection command to the \injectorAgent{} on the proper Compute node with the target tap device.
    

\noindent
$\rhd$ \textbf{Internal virtual router interfaces}. \toolname{} can inject into traffic over the virtual router interfaces that face a private virtual network of the tenant. It targets the port identified as {\lmttfont network:router\_interface}, which represents a network interface on the Network node ({\lmttfont qr-XXX}) connected to a virtual router through the \emph{integration} bridge {\lmttfont br-int} (left side of \figurename{}~\ref{fig:openstack_networking}). In turn, the virtual router has a dedicated network Linux namespace, and uses routing tables and iptables rules to exchange traffic between virtual subnets. The \toolname{} \netcastAgent{} matches the target virtual router interface with this network interface, and sends an injection command to the \injectorAgent{} on the Network node.

\noindent
$\rhd$ \textbf{External virtual router interfaces}. Similar to the previous item, \toolname{} targets ports identified as {\lmttfont network:router\_gateway} to inject faults into virtual router interfaces towards public networks. The port represents a network interface on the Network node (like {\lmttfont qg-XXX}, which is connected to the \emph{external} bridge {\lmttfont br-ex} (left side of \figurename{}~\ref{fig:openstack_networking}). To inject there, the \toolname{} \netcastAgent{} triggers the \injectorAgent{} on that Network node.

\noindent
$\rhd$ \textbf{Floating IPs}. These addresses are like publicly routable IPs that one typically buys from an ISP. \toolname{} targets ports identified as {\lmttfont network:floatingip} to inject faults into traffic flowing through a specific floating IP. The port represents a network interface ({\lmttfont qg-XXX}) connected to the \emph{external} bridge {\lmttfont br-ex}. This network interface is assigned to a specific IP subnet handled by a virtual router. The \injectorAgent{} on the Network node targets this interface for injecting into floating IPs.


\subsection{Injection of Virtual Configuration Faults}


In addition to network traffic faults, \toolname{} allows injecting \emph{configuration faults} that impact the availability of virtual network resources. This injection emulates configuration faults in which a cloud architect accidentally removes or misconfigures a network element in a production cloud environment. \toolname{} is designed to delete OpenStack network resources in a controlled and safe way, such that the resources can be automatically restored after a fault injection experiment. 
Our implementation of \toolname{} supports the deletion of networks, routers, floating IPs, and ports, by providing \textit{restore} REST APIs. In turn, these APIs use the {\lmttfont list}, {\lmttfont update}, and {\lmttfont remove} OpenStack REST APIs \cite{openstack_networking_API}. The effect of deletion is removing OpenStack network ports, along with networking resources on the hosting node. To restore them, \toolname{} saves the configuration of deleted resources: for example, when deleting a virtual NIC, it saves the instance ID, the subnet ID, the network ID, and any associated floating IPs, and reconfigures them once the virtual NIC is brought up again.



\subsection{Workload Generators}

\toolname{} provides REST APIs to run pre-defined or custom workload generators during the fault injection experiments. To this purpose, \toolname{} uses OpenStack Heat, the orchestration service that enables the deployment of instances, volumes, and other OpenStack services using YAML-based templates. As pre-defined workloads, \toolname{} includes two popular traffic generators, namely \textit{Iperf} \cite{iperf}, which is a TCP/UDP network bandwidth measurement tool, and \textit{Apache JMeter} \cite{jmeter}, which is a load testing tool for analyzing and measuring the performance of a variety of services, with a focus on web applications.

\toolname{} includes YAML templates and pre-built VM images, to automatically deploy workload generators across the virtual network. In the case of \textit{Iperf}, the user can select where to attach two instances (a client and a server) in the virtual network, such that they generate TCP or UDP traffic flows between two points of the virtual network. In the case of \textit{JMeter}, the user can attach an instance (a web client) into a selected point of the network, to generate traffic towards a web service inside the virtual network. For both workload generators, the user can provide configuration for the volume and type of traffic. Moreover, \toolname{} allows the user to integrate their own workload generator, by implementing hook functions to be called when the workload has to be started or stopped.



\section{Case Studies and Experimental Results}
\label{sec:casestudy}

In this section, we analyze two complex scenarios of cloud-based applications, namely (i) an NFV IP Multimedia Subsystem (IMS), and (ii) a three-tier, fault-tolerant web application. 
In the first case study, we apply \toolname{} for \textbf{extending the functional tests of the IMS with fault injection}. In the second case study, we apply \toolname{} for the \textbf{redundancy and capacity planning} of a three-tier web application on the cloud. 
The experiments were performed on a private cloud, using OpenStack version {\lmttfont 3.12.2} (release Pike). 
The hardware configuration of the machines in the testbed includes: 32 virtual Intel Xeon CPUs (E5-2630L v3 @ 1.80GHz); 64GB RAM; 1TB storage; Linux Ubuntu 18.04.2 LTS.

\subsection{NFV IP Multimedia Subsystem (IMS)}
\label{subsec:ims_case_study}

This section shows how \toolname{} can complement functional tests of a complex distributed system, and which insights can be gained from fault injection using the proposed solution. We targeted VNF software from the \emph{Clearwater} project \cite{clearwater_old,clearwater_ims_core} ({\lmttfont release-123}), which is a popular open-source,  cloud-native implementation of an IMS. The main components of Clearwater are:

\begin{itemize}

	\item \textbf{{\lmttfont Bono}}: The SIP edge proxy, which provides both SIP IMS Gm and WebRTC interfaces to clients. The Bono node is the first point of contact between clients and the IMS. Clients are linked to a specific Bono node at registration.
	
	\item \textbf{{\lmttfont Sprout}}: The SIP registrar and authoritative routing proxy, which handles client authentication, and provides the bulk of I-CSCF and S-CSCF functions. According to the SIP protocol about transactions and dialogs \cite{rosenberg2002rfc3261}, the Sprout nodes are transaction-stateful but not dialog-stateful: if a request flows through a specific Sprout, the corresponding response(s) must also flow through that Sprout.
	
	\item \textbf{{\lmttfont Dime}}: The \textit{Diameter gateway}, which run \textit{Homestead} and \textit{Ralf} components. Homestead provides a web service interface to Sprout for retrieving authentication credentials and user profile information. It can either use a local master of the data, or it can pull the data from an IMS-compliant Home Subscriber Service (HSS). The Homestead node works as an HSS cache, it is stateless and handles authentication credentials and user profiles by using data stored on the {\lmttfont Vellum} node. Ralf is used for reporting billable events.
	
	\item \textbf{{\lmttfont Vellum}}: It is used to maintain all long-lived states in the IMS through distributed storage. In particular, Vellum leverages \emph{Cassandra} \cite{lakshman2010cassandra} to store authentication credentials and profile information, \emph{Chronos} to provide reliable timer service, and \emph{Memcached} for storing registration and session state.
	
	\item \textbf{{\lmttfont Homer}}: XML Document Management Server that stores MMTEL service settings, using a Cassandra datastore running on the Vellum node.
	
\end{itemize}

These VNFs form a \textit{service function chain} (SFC) that provides the IMS services. In addition to these VNFs, we set up the following nodes in the testbed:

\begin{itemize}

    \item \textbf{{\lmttfont Ellis}}: It provides self sign-up, password management, line management and control of MMTEL service settings. We use this node to run the functional test suite provided by the Clearwater \cite{clearwater_live_test}.
    
    \item \textbf{{\lmttfont DNS Server}}: It provides a name resolution service for the entire IMS. We use {\lmttfont bind} as an implementation of the DNS server \cite{bind_tool}.
    
    \item \textbf{{\lmttfont Load Balancer}}: A node that forwards and balances IMS requests to the {\lmttfont Bono} VNFs. We use {\lmttfont haproxy} as load balancer \cite{haproxy_tool}.
    
\end{itemize}

\begin{figure}[!h]
  \centering
  \includegraphics[width=\columnwidth]{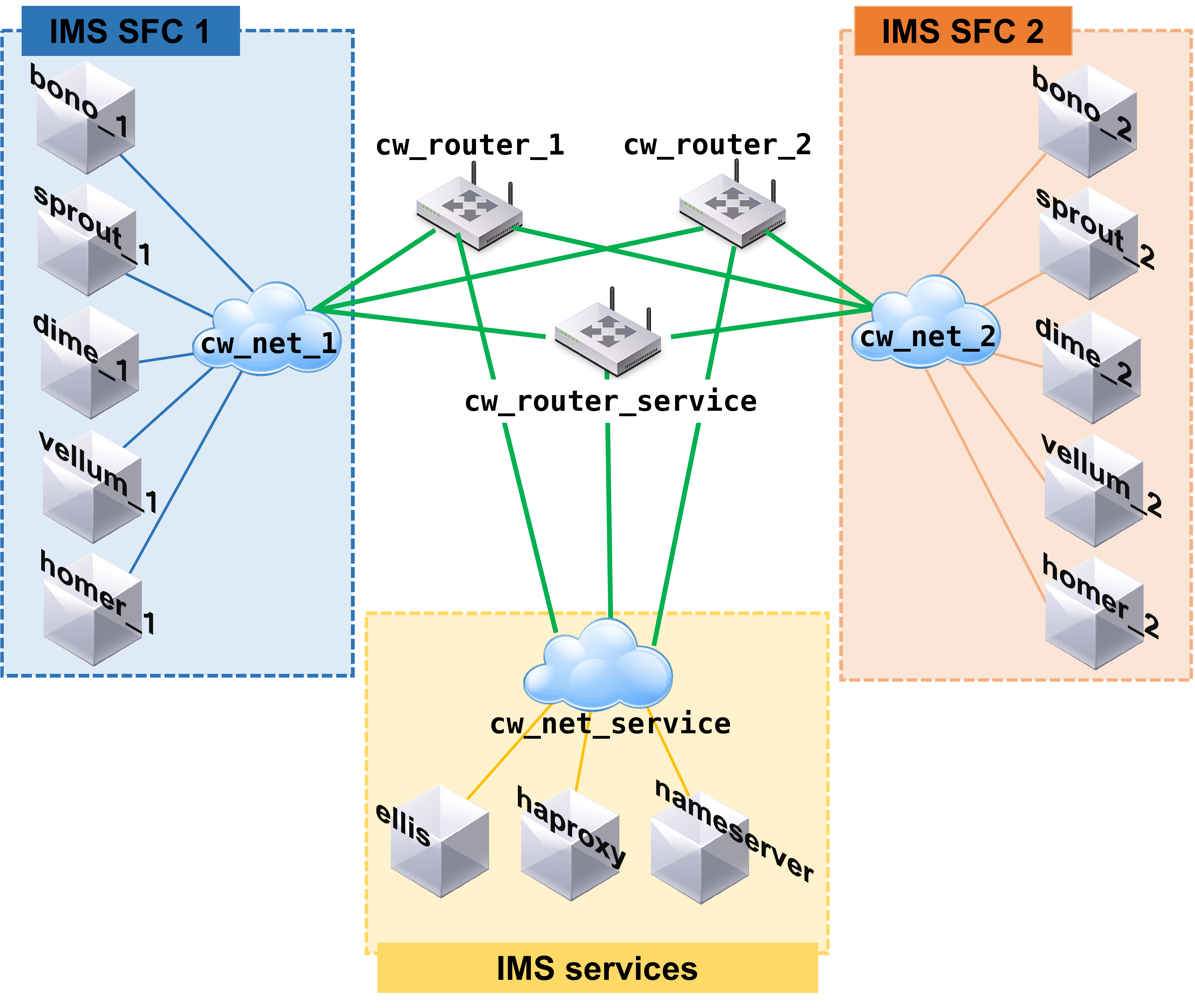}
  \caption{Virtual network topology of the IMS.}
  \label{fig:ims_testbed}
\end{figure}

We deploy the Clearwater IMS via OpenStack using a dual-redundant SFC, in which all the VNFs are replicated with an active/active schema, and each SFC is running on separated virtual network segments. This kind of deployment is typical of a high-available IMS, since it can mask single point of failures caused by network faults.  \figurename{}~\ref{fig:ims_testbed} shows the topology of the virtual network. The network segments {\lmttfont cw\_net}, {\lmttfont cw\_net\_2}, and {\lmttfont cw\_net\_service} are attached, respectively, to the virtual routers {\lmttfont cw\_router}, {\lmttfont cw\_router\_2}, and {\lmttfont cw\_router\_service}, which route network packets across the network segments.

We used \toolname{} to perform experiments on Clearwater IMS, by injecting faults during execution of the functional test suite \textit{Clearwater-live-test} \cite{clearwater_live_test}. Clearwater's developers recommend using this suite when creating a new IMS deployment, in order to get confirmation that the IMS is properly installed, configured, and working. The functional tests include basic calls, call diversion (forwarding), subscribing events, Globally Routable User Agent URIs (GRUUs) calls, and others. We refer the reader to the \textit{Clearwater-live-test} documentation for the full list of tests \cite{clearwater_live_test}. 

We performed two groups of fault injection test plans, where faults are injected respectively on (i) one of the two SFC network segments, and (ii) one of the two {\lmttfont Sprout} VNF instances (we reported tests only for the Sprout VNF both for the sake of brevity and since it is the most critical VNF in the IMS SFC). During the execution of each test case, we simulate the occurrence of a network fault in one part of the network (i.e., a network segment, or a {\lmttfont Sprout} instance), and let the other part continue the execution with any injected fault. Clearwater has been designed to be highly available through horizontal scalability, and configured by us with a dual-redundant schema. Therefore, the expected, proper behavior is that the functional tests must still eventually succeed even if one part of the network fails. Thus, failures will denote fault management issues in Clearwater.

An experiment executes one functional test case repeatedly for $20$ times for statistical significance purposes. The early repetitions are performed with no fault injection, in order to warm-up the target system into a steady-state before the injection \cite{vieira2007comparing,vieira2007benchmarking,kanoun2008dependability}. We configured \toolname{} to trigger fault injection in the middle of the repetitions, that is, after $10$ seconds since the beginning of the experiment. 
Moreover, we configured \toolname{} to emulate persistent faults in one part of the network, i.e., all network packets that flow through the target resource (network segment or {\lmttfont Sprout} instance) are injected with a fault. The IMS is expected to eventually recover from these faults, by handling the requests through the remaining available replicas of the target resource.

We performed experiments with different types of faults, by injecting packet losses, delays, and corruption faults. 
We repeated this process for $47$ test cases provided by the \textit{Clearwater-live-test} suite, for each of the three fault types (loss, delay, corruption), and target (segment, {\lmttfont Sprout} VNF), for a total of $282$ fault injection experiments.


\begin{table}[!htb]
\centering
\caption{Distribution of failed tests within the fault injection test plan on one of the redundant network segments.}
\label{table:ims_network_faults}
\resizebox{\columnwidth}{!}{
\sffamily 
\normalsize	
\rowcolors{2}{gray!15}{white}
\begin{tabular}{>{\centering\arraybackslash}c>{\centering\arraybackslash}c>{\centering\arraybackslash}c>{\centering\arraybackslash}c}

\cmidrule{2-4}
\multicolumn{1}{c}{} & \multicolumn{3}{c}{\textbf{Fault}}\\
\rowcolor{LightCyan}
\toprule\toprule
 \textbf{Test Category} & \textbf{Corruption} & \textbf{Delay} & \textbf{Loss} \\ \toprule\toprule
Basic Call                                                                    & 0/7                                   & 0/7                                   & 0/7                                   \\ \hline
Basic Registration                                                            & 1/1                                   & 0/1                                   & 0/1                                   \\ \hline
CANCEL                                                                        & 0/1                                   & 0/1                                   & 0/1                                   \\ \hline
Call Barring                                                                  & 0/1                                   & 0/1                                   & 0/1                                   \\ \hline
Call Diversion                                                                & 1/11                                  & 0/11                                  & 2/11                                  \\ \hline
Call Forking                                                                  & 0/3                                   & 1/3                                   & 0/3                                   \\ \hline
Call Waiting                                                                  & 1/2                                   & 0/2                                   & 0/2                                   \\ \hline
Filtering                                                                     & 0/6                                   & 0/6                                   & 1/6                                   \\ \hline
GRUU Call                                                                     & 1/10                                  & 0/10                                  & 0/10                                  \\ \hline
GRUU REGISTER                                                                 & 0/3                                   & 1/3                                   & 0/3                                   \\ \hline
SUBSCRIBE                                                                     & 0/2                                   & 0/2                                   & 1/2           \\ \hline\hline
\rowcolor{lightgray} \textbf{Total}          & \pmb{$4/47 (9\%)$}                & \pmb{$2/47 (4\%)$}              & \pmb{$4/47 (9\%)$}           \\ \hline\hline
\end{tabular}
}
\end{table}

\begin{table}[!htb]
\centering
\caption{Distribution of failed tests within the fault injection test plan on an Sprout VNF instance.}
\label{table:ims_sprout_faults}
\resizebox{\columnwidth}{!}{
\sffamily 
\normalsize
\rowcolors{2}{gray!15}{white}
\begin{tabular}{>{\centering\arraybackslash}c>{\centering\arraybackslash}c>{\centering\arraybackslash}c>{\centering\arraybackslash}c}

\cmidrule{2-4}
\multicolumn{1}{c}{} & \multicolumn{3}{c}{\textbf{Fault}}\\
\rowcolor{LightCyan}
\toprule\toprule
 \textbf{Test Category} & \textbf{Corruption} & \textbf{Delay} & \textbf{Loss} \\ \toprule\toprule
Basic Call                                                                    & 1/7                                   & 3/7                                   & 2/7                                   \\ \hline
Basic Registration                                                            & 1/1                                   & 1/1                                   & 0/1                                   \\ \hline
CANCEL                                                                        & 0/1                                   & 0/1                                   & 0/1                                   \\ \hline
Call Barring                                                                  & 0/1                                   & 0/1                                   & 0/1                                   \\ \hline
Call Diversion                                                                & 0/11                                  & 3/11                                  & 1/11                                  \\ \hline
Call Forking                                                                  & 0/3                                   & 0/3                                   & 0/3                                   \\ \hline
Call Waiting                                                                  & 0/2                                   & 0/2                                   & 0/2                                   \\ \hline
Filtering                                                                     & 0/6                                   & 4/6                                   & 0/6                                   \\ \hline
GRUU Call                                                                     & 0/10                                  & 1/10                                  & 1/10                                  \\ \hline
GRUU REGISTER                                                                 & 2/3                                   & 3/3                                   & 1/3                                   \\ \hline
SUBSCRIBE                                                                     & 0/2                                   & 1/2                                   & 0/2     \\\hline\hline
\rowcolor{lightgray} \textbf{Total}          & \pmb{$4/47 (9\%)$}                & \pmb{$16/47 (34\%)$}              & \pmb{$5/47 (11\%)$}           \\ \hline\hline
\end{tabular}
}
\end{table}




Despite the fault-tolerant configuration of the Clearwater IMS, we found that the injected faults have a non-negligible impact on the IMS, both in the case of injections in the network segment, and of injections in the {\lmttfont Sprout} instance. 
Table~\ref{table:ims_network_faults} and Table~\ref{table:ims_sprout_faults} show the distributions of failures in the two fault injection test plans. In particular, the tables count \emph{how many test cases experienced more than one failure} (i.e., termination with a \textit{failed} status) across repetitions of the test case. For example, the tables show that there were 7 test cases in the \emph{Basic Call} category. When injecting corruptions, one of these 7 test cases failed at least one time across the $20$ repetitions performed during the experiment against Sprout VNF.

\begin{table*}[!htb]
\centering
\caption{Most frequent errors reported by the Sprout VNF during execution of the fault injection test plan.}
\label{table:sprout_injections_details}
\resizebox{2\columnwidth}{!}{
\sffamily 
\footnotesize
\begin{tabular}{L{3.2cm}L{3.8cm}L{5cm}L{6.7cm}}

\toprule \toprule
\rowcolor{LightCyan}
\normalsize\textbf{Component} & \normalsize\textbf{Error Message} & \normalsize\textbf{Reason} & \normalsize\textbf{Reference}  \\

\toprule \toprule

 & Could not get subscriber data from HSS & Either not found the subscriber on the HSS, or unable to communicate with the HSS   successfully. & \href{https://github.com/Metaswitch/sprout/blob/release-123/src/hssconnection.cpp\#L713}{https://github.com/Metaswitch/sprout/blob/release-123/src/hssconnection.cpp\#L713}\\ 

 \rowcolor{lightgray!50} \cellcolor{white} \multirow{-4}{*}{{\lmttfont hssconnection.cpp}} & Failed to get Authentication Vector for \textit{NUMBER}@example.com                             & Failed to retrieve a JSON object from a path on the HSS server regarding function to get an Authentication Vector as JSON object. & \href{https://github.com/Metaswitch/sprout/blob/release-123/src/hssconnection.cpp\#L148}{https://github.com/Metaswitch/sprout/blob/release-123/src/hssconnection.cpp\#L148}\\ \midrule \midrule

    & cURL failure with cURL error code 28   & HSS responds with an {\lmttfont HTTP 503 Service Unavailable} due to timeouts (curl error 28).  & \href{https://github.com/Metaswitch/cpp-common/blob/release-123/src/httpclient.cpp\#L712}{https://github.com/Metaswitch/cpp-common/blob/release-123/src/httpclient.cpp\#L712}                                    \\

    \rowcolor{lightgray!50} \cellcolor{white} \multirow{-4}{*}{{\lmttfont httpclient.cpp}} & cURL failure with cURL error code 0 & HSS responds with an {\lmttfont HTTP 500 Server error } (curl error 0 is ignored). & \href{https://github.com/Metaswitch/cpp-common/blob/release-123/src/httpclient.cpp\#L712}{https://github.com/Metaswitch/cpp-common/blob/release-123/src/httpclient.cpp\#L712}     \\ \midrule \midrule


 & 
    Rejecting REGISTER request following error communicating with the HSS & 
    This is either a server error on the HSS, or a error decoding the response & \href{https://github.com/Metaswitch/sprout/blob/release-123/src/hss\_sip\_mapping.cpp\#L90}{https://github.com/Metaswitch/sprout/blob/release-123/src/hss\_sip\_mapping.cpp\#L90}            \\ 
    
    \rowcolor{lightgray!50} \cellcolor{white} \multirow{4}{*}{{\lmttfont hss\_sip\_mapping.cpp}} & 
    Rejecting REGISTER request as unable to contact HSS &
    The HSS is unavailable. The client should retry on timeout but no other Clearwater nodes should (as Sprout will already have retried on timeout). Reject with a 504 (503 is used for overload). & \href{https://github.com/Metaswitch/sprout/blob/release-123/src/hss\_sip\_mapping.cpp\#L81}{https://github.com/Metaswitch/sprout/blob/release-123/src/hss\_sip\_mapping.cpp\#L81}            \\ 

     &
    Rejecting SUBSCRIBE request following failure to register on the HSS  &
    Failed to register a subscriber at the HSS. This may be because the HSS is unavailable, the public identity doesn't exist, the public identity doesn't belong to the private identity, or there was an error communicating with the HSS. In particular, it is successfully contacted Homestead, but the user wasn't registered. 
    &   \href{https://github.com/Metaswitch/sprout/blob/release-123/src/hss\_sip\_mapping.cpp\#L60}{https://github.com/Metaswitch/sprout/blob/release-123/src/hss\_sip\_mapping.cpp\#L60}            \\ \midrule \midrule

    \rowcolor{lightgray!50} \cellcolor{white} {\lmttfont registrarsproutlet.cpp} &
    Failed to get AoR binding for sip:\textit{NUMBER}@example.com from store & 
    Failed to get data for the Address of Record (AOR) because there is no connection to the store. & 
    \href{https://github.com/Metaswitch/sprout/blob/release-123/src/registrarsproutlet.cpp\#L852}{https://github.com/Metaswitch/sprout/blob/release-123/src/registrarsproutlet.cpp\#L852} \\
    \bottomrule \bottomrule

\end{tabular}
}
\end{table*}

Between $4\%$ and $34\%$ of the test cases (depending on the injected fault), the IMS did not handle the injected fault, since there were multiple repetitions of the test case terminated with a failure. 
This result points out that even if the IMS is configured with redundant VNFs, faults can still lead to failures experienced by users. This behavior is caused by a design issue in Clearwater, since it does not re-balance all the IMS requests upon failure (e.g., when a request times out, the destination should not be contacted again). 
Furthermore, the results show that all the injected fault types (i.e., corruption, delay, and loss) have an impact on the IMS, since faults in the network segment cause failures in at least $4\%$ of cases (Table~\ref{table:ims_network_faults}), and faults in the Sprout VNF traffic cause failures in at least $9\%$ of cases (Table~\ref{table:ims_sprout_faults}). 

Failures are more likely for faults in the Sprout VNF traffic, as there is a higher number of failed tests compared to the injections in the network segment. 
This result is explained by the importance of the Sprout VNF in the IMS service chain, since it is pivotal for providing core functionalities about Call Session Control Function (CSCF), such as SIP registrations, forwarding SIP messages to the specific application server, and others. Once a session is routed towards the network segment with the faulty Sprout VNF, it is unable to recover using the other network segment. Instead, in some cases, when one of the network segments is unavailable, the IMS is still able to serve the session using the other segment.

In both the fault injection test plans on the network segment and on the Sprout VNF, we observed several errors reported in the logs by the Sprout VNF. Table~\ref{table:sprout_injections_details} shows detailed information about the most frequent errors, including the error message, the reason for the error according to our analysis of the Sprout source code, and a reference to the Sprout source code. 
In particular, there is a high amount of errors related to interactions with Homestead VNF nodes. Since the Homestead VNF acts as an HSS cache, the Sprout VNF interacts with Homestead to read subscribers' data, such as IMS subscriptions and AKA authentication vectors. 

Information about error sites in the source code (Table~\ref{table:sprout_injections_details}) is useful for IMS developers to introduce more error handling code for the most critical interactions, such as, by adding mechanisms for detecting failures through timeouts and for retrying failed requests. In the case of Sprout node failures, the session should be switched to a different Sprout node, by using the Route header to specify a different cluster, so that P-CSCF requests are routed to an available replica.



Finally, we found that injections have also an impact on the fault-free VNFs, where we observed error messages with high severity. Despite these VNFs execute in the non-injected network segment, they still experience effects propagated from the injected network segment. This counterintuitive behavior is explained by the interactions between VNFs located on different network segments. These interactions occur via DNS load balancing, which distributes IMS requests between all the segments, regardless of whether the node is currently available or not. 
A possible countermeasure is to introduce a mechanism that periodically updates the configuration of load balancing according to the current state of the VNFs.

\revision{\toolname{} has been effective at automating these experiments on the IMS and to obtaining these insights. Differently from many fault injection tools, \toolname{} integrates automation mechanisms for deploying and for running the system several times, and for collecting and analyzing log data. These features have been useful to amplify an existing test suite with fault injection, without the need for intrusive changes in the virtual machines.}

\subsection{Multi-tier Cloud Web Application}
\label{subsec:netcast_demo_multitier}

This section presents an experimental case study on how \toolname{} can be applied for supporting capacity and redundancy planning. The system designer can use fault injection to assess whether the system achieves high performance even in the case of failures. Therefore, the system designer can identify bottleneck resources, and deploy more resources as needed.

\begin{figure}[!h]
	\centering
	\includegraphics[width=\columnwidth]{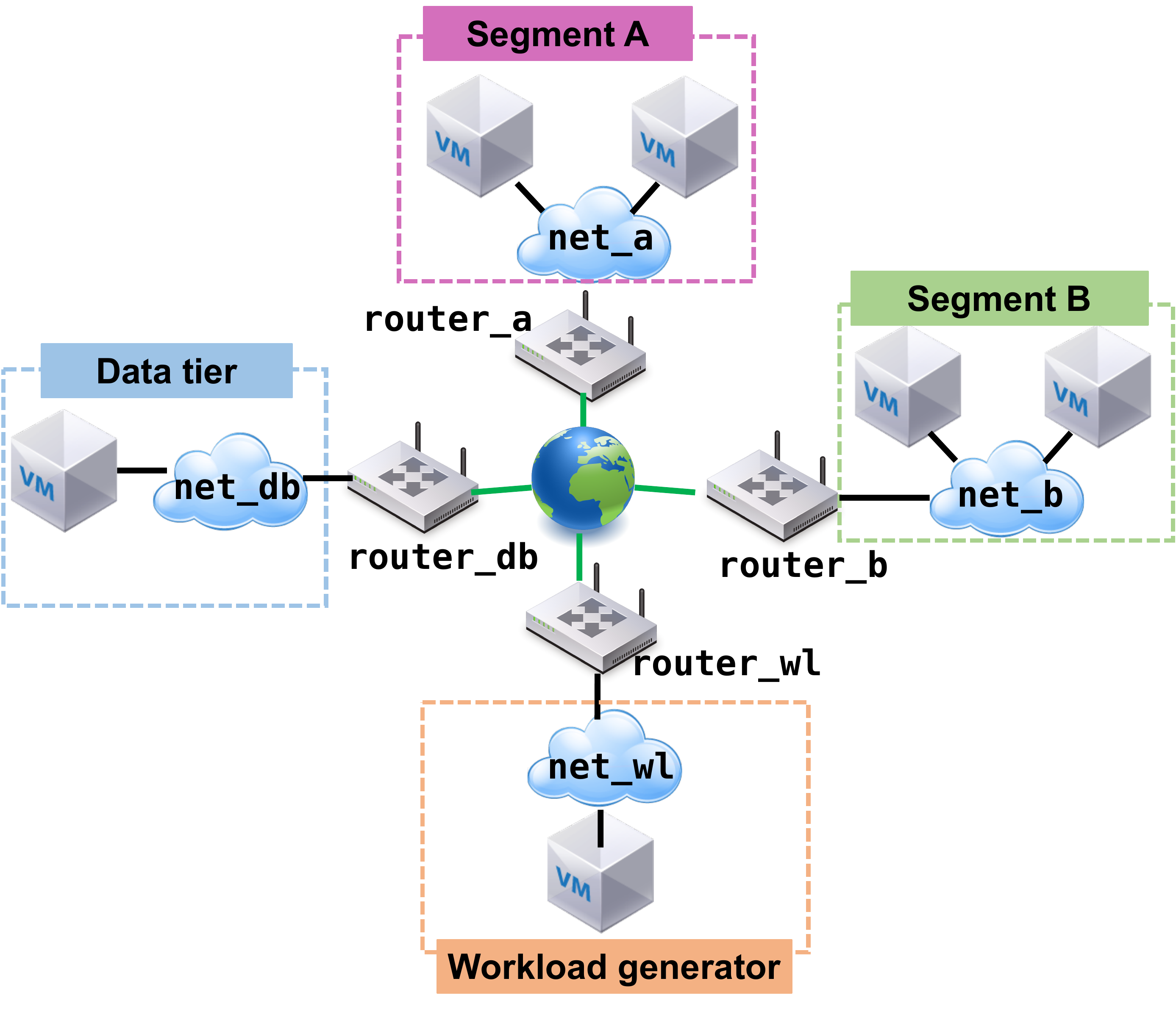}
	\caption{Virtual network topology of the multi-tier cloud web application.}
	\label{fig:architecture_3tier}
\end{figure}

We performed fault injection experiments on a dual-redundant virtual network, where we deploy a 3-tier web application. The goal of the experiments is to evaluate whether the resources allocated to the virtual network and the virtual machines are sufficient to achieve high performance despite failures. 
The web application is a CRUD one (Create-Read-Update-Delete), which consists of web pages to get information about the user's account and personal data, to update user information, to perform data entry into the database, and other operations such as data deletion and user management.

The web application is deployed with a high-availability schema using OpenStack, as showed in Figure~\ref{fig:architecture_3tier}. It consists of two replicated network segments. For each segment, we run two virtual machines, one acting as \textit{presentation tier}, e.g., to serve web resources, such as HTML and images, and the other acting as \textit{application tier}, i.e., implementing the business logic, using the PHP language. 
Moreover, we run a virtual machine to act as \textit{data tier}, i.e., to provide data persistence, which runs the MySQL DBMS. 
The VMs are equipped with 2 vCPUs, 1 GB RAM and 3 GB of HDD, and run Linux Ubuntu 16.04.5 LTS (Xenial).
We used \toolname{} to deploy an additional virtual machine to act as \textit{workload generator}, by using the JMeter application \cite{jmeter} to generate HTTP requests for the web application. 
We configured JMeter to simulate $25$ concurrent users, which invoke the web application to perform a randomly-selected operation, with a load of $240$ reqs/minute.

The virtual network is based on the OpenStack Load Balancer as a Service (LBaaS), an HAProxy-based load balancer \cite{openstack_lbaas}. We use OpenStack LBaaS to balance the load among the two replicated network segments A and B (Figure~\ref{fig:architecture_3tier}). 
We configured the load balancer to use the round-robin algorithm, for equal balancing of requests between the segments. Moreover, we enabled the OpenStack LBaaS built-in \textit{health monitor}, which keeps track of the availability of the two network segments and, if one of them is unresponsive, diverts traffic away from the \textit{failed} segment. We configure the health monitor with an integrity check period equal to 5 seconds, maximum connection retry equal to 3 seconds, and timeout of health checks equal to 5 seconds.


We adopted \toolname{} to perform fault injection, by targeting one of the two replicated network segments (segment B), and the network segment of the data tier. In the case of the replicated network segment, we injected both on the traffic of the entire segment, and on the floating IP of the application tier server.
We performed distinct experiments to inject different types of faults, including packet loss, packet corruption, and packet delay, at different levels of intensity ($25\%$, $50\%$, and $75\%$, rate of packets corrupted or lost, and $500$, $1000$, and $1500$ ms of packet delay).

\begin{table*}[!h]
\centering
\caption{Results from fault injection experiments on the multi-tier cloud web application.}
\label{tab:fault_injection_test_plan}
\sffamily 
\footnotesize
\resizebox{2\columnwidth}{!}{
\rowcolors{2}{gray!15}{white}
\begin{tabular}{ccccccc}
\toprule\toprule
\rowcolor{LightCyan} 
\textbf{Exp. ID} & \textbf{Fault Type} & \textbf{Intensity} & \textbf{Target} & \textbf{Throughput} & \textbf{Response Time} & \textbf{Latency} \\ \toprule\toprule
0 & No fault injected & N/A & N/A & 3.80 req/s & 2558 ms & 2027 ms \\ \hline

1 & Packet Loss & 25\% & App. Network Segment & \textbf{1.12 req/s} & \textbf{7481 ms} & \textbf{5327 ms} \\ \hline
2 & Packet Loss & 50\% & App. Network Segment & \textbf{2.51 req/s} & \textbf{5913 ms} & \textbf{4625 ms} \\ \hline
3 & Packet Loss & 75\% & App. Network Segment & 3.45 req/s & 2989 ms & 2469 ms \\ \hline

4 & Packet Corruption & 25\% & App. Server Traffic & 3.65 req/s & 2640 ms & 2028 ms \\ \hline
5 & Packet Corruption & 50\% & App. Server Traffic & 3.49 req/s & 3442 ms & 2659 ms \\ \hline
6 & Packet Corruption & 75\% & App. Server Traffic & 3.23 req/s & \textbf{4259 ms} & 3592 ms \\ \hline

7 & Packet Loss & 25\% & Data Network Segment & 3.43 req/s & 3158 ms & 2254 ms \\ \hline
8 & Packet Loss & 50\% & Data Network Segment & \textbf{0.83 req/s} & \textbf{4108 ms} & 3303 ms \\ \hline
9 & Packet Loss & 75\% & Data Network Segment & \textbf{1.11 req/s} & 3899 ms & 3074 ms \\ \hline

10 & Packet Delay & 500ms & Data Network Segment & 3.26 req/s & 3439 ms & 2632 ms \\ \hline
11 & Packet Delay & 1000ms & Data Network Segment & 3.05 req/s & \textbf{4074 ms} & 3235 ms \\ \hline
12 & Packet Delay & 1500ms & Data Network Segment & \textbf{2.80 req/s} & \textbf{4702 ms} & 3936 ms \\ \bottomrule

\end{tabular}%
}
\end{table*}

\begin{figure*}[!h]
	\centering
	\includegraphics[width=0.7\linewidth]{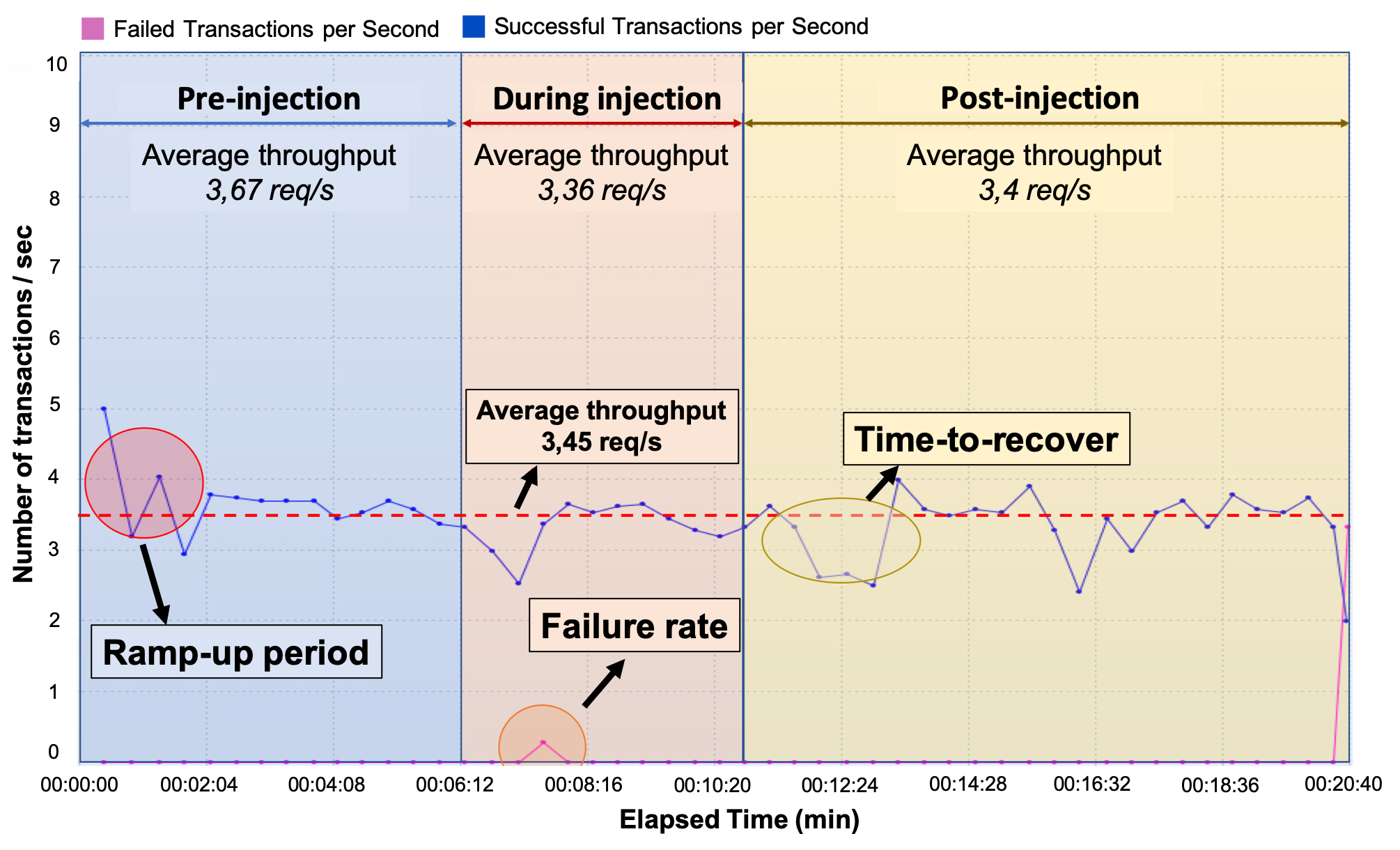}
	\caption{Throughput and failure rate during packet loss injection on a virtual network segment.}
	\label{fig:packetloss_transactions}
\end{figure*}

Table~\ref{tab:fault_injection_test_plan} reports on a selection of experiments that are representative of the failures of the cloud web application. 
For each experiment, the table shows performance metrics as perceived by clients, i.e., \emph{throughput}, \emph{response time} (i.e., the difference between the time when a request was sent and the time when the response has been fully received), and \emph{latency} (i.e., the difference between the time when a request was sent and the time when the response has started to be received). 
As a reference for the interpretation of the results, the first row in the table shows the average throughput, latency, and response time from executions with no fault injection, and one additional row for each fault injection experiment. The largest deviations from the fault-free run are emphasized in bold. 
The main results include:

\begin{itemize}

\item \textbf{Packet loss in the application tier network segment traffic (Exp. 1, 2, 3)}. These experiments target the virtual network of the B segment within the web application. We can notice that, even if the packet loss is relatively low (25\%), there is still a high-performance degradation concerning throughput, response time, and latency. On the contrary, with a higher percentage of packet loss (i.e., 50\% and 75\%), the performance degradation was less severe, since the load balancer has been able to switch most of the traffic on the other virtual network (i.e., segment A). 
Figure~\ref{fig:packetloss_transactions} plots the data about one of these experiments (75\% of packet loss). The figure highlight the three phases of the experiment, i.e., measurements collected before fault injection, during fault injection, and after fault injection. During injection, the throughput degrades to around $2$ req/s, and there are some failed HTTP transactions. 
In the meanwhile, the health manager initially tries to contact the faulty virtual network segment for 15 seconds (3 attempts of 5 seconds each). During this period, packets continue to get lost and therefore there is an increase in failure rate. After 15 seconds, the load balancer performs a switch on the other (healthy) network segment, thus isolating the faulty segment. After the failure is detected and managed, the normal throughput is restored. 
This experiment demonstrates the ability of the high-availability mechanisms to recover as intended. 
In the post-injection phase, when the fault injector is disabled, there is a period in which the throughput is again low, since traffic is routed again towards the injected network segment (now healthy), which takes some time for warm-up to recover the original performance levels. 

\item \textbf{Packet corruption in the application tier server traffic (Exp. 4, 5, 6)}: These experiments target the traffic that flows through the floating IP of the VM server in one of the network segments with the application tier. The web application experiences a higher response time and latency at the higher rate of packet corruption (75\%). In this experiment, the load balancer was still able to switch traffic on the other virtual network segment, thus mitigating the effects of the faults on the throughput.

\item \textbf{Packet loss in the data tier network segment traffic (Exp. 7, 8, 9)}: These experiments target the virtual network segment of the data tier. Since this part of the web application is not replicated, the injected faults cause a significant performance slowdown, with very low throughput and a high response time. The slowdown is exacerbated at higher rates of packet loss.


\item \textbf{Packet delay in the data tier network segment traffic (Exp. 10, 11, 12)}: These experiments inject delay faults with increasing severity (500 ms, 1000 ms, and 1500 ms of packet delay) on the virtual network segment of the data tier. These injected delays propagate throughout the tiers, causing the degradation of the throughput, and the increase of response time and latency.


\end{itemize}

Fault injection experiments provide useful feedback about capacity and redundancy planning.
One of the outcomes of the experiments is measurements for the \emph{time-to-recovery}, both during the fault injection phase and during the post-injection period. From these measurements, system designers can learn about the ability of their system to restore a steady-state both when a fault occurs, and when the fault is removed from the system. 
These measurements allow them to assess whether recovery is quick enough to satisfy service level requirements (e.g., ``five-nines'' SLAs, meaning that the system can only afford few minutes of yearly outages). If the requirements are not fulfilled, the measurements provide a quantitative goal to optimize for, by supporting system designers at tuning their failure management mechanisms (e.g., heartbeat periodicity and timeout).

In some cases (in particular, experiments 9 and 10), redundant network segments do not guarantee high availability. Despite redundancy, the failure of only one network segment is still able to significantly reduce throughput and increase response time. This behavior was caused by the low intensity of the injected faults, which were not enough frequent to trigger the health manager, but they still affected performance perceived by the clients. 
These problems are referred to as \emph{gray} failures, that is, failures whose manifestations are subtle and defy detection, such as random packet loss and performance degradation of network devices due to wear-out. These failures are today one of the most important and problematic causes of incidents in cloud infrastructures \cite{huang2017gray,gunawi2018fail}.
Due to their subtle effects, these failures require special failure management mechanisms to be handled, such as, by performing deeper monitoring of performance and resource metrics to detect and to isolate the faulty elements \cite{guo2015pingmesh,gill2011understanding}.

The network segment of the data tier turned out to be another vulnerable point of the virtual network. Unfortunately, introducing redundancy in the data tier is a difficult challenge, since relational DBMSs can only be replicated using sophisticated mechanisms that come with a high cost in terms of lower performance \cite{ozsu1999principles}. If replication is not feasible, the system designer should at least make the network layer more reliable (e.g., by deploying on data centers with redundant physical connections); otherwise, they should consider redesigning the application to use NoSQL datastores, which are more suitable for horizontal elasticity and replication \cite{klein2015performance,ventura2016experimental,cotroneo2020dependability}. Therefore, system designers need to make design choices to achieve good trade-offs between performance and costs.
\revision{
This case studies confirms that \toolname{} is useful for designers to make such decisions: it integrates with cloud management software that designers already use for managing their virtual networks, thus better supporting them at configuring a fault injection plan without the need for using external tools; moreover, \toolname{} integrates log collection and analysis of availability and performance, which is useful to get quantitative measures for making design decisions.
}

\section{Qualitative Evaluation}
\label{sec:discussion}

Performing fault injection for cloud applications poses several challenges. The first is represented by the high cost (e.g., in terms of developers' time) for automating fault injection, since cloud applications typically involve many with several virtualized network resources, including virtual NICs, switches, routers, load balancers, interconnected by complex virtual topologies. The underlying physical infrastructure is even more complex: for example, large OpenStack cloud deployments typically have hundreds of network and compute nodes, as well as tens of controller nodes \cite{openstack_deployments}. Moreover, fault injection must assure that the injections on a cloud tenant do not affect other tenants that share the same infrastructure. 
In this section, we discuss how the proposed solution alleviates these issues and the portability of the solution across cloud management platforms.

\vspace{3pt}
\noindent
$\rhd$ \textbf{Cost-efficiency of fault injection.} \toolname{} can automatically orchestrate fault injection experiments, even for large virtual networks. It provides both REST APIs and a visual dashboard that hide the complexity of the virtual networks and of the underlying data center infrastructure, and reduce the efforts needed to configure and to run the experiments. In order to qualitatively evaluate the gain of using \toolname{} in terms of saved developers' efforts, we estimate the efforts that would be needed to implement fault injection in an OpenStack-based cloud without using \toolname{}. We consider the hypothetical case of a small network with only a client and a server, where fault injection is automated using custom-made Python programs based on OpenStack APIs, to be written by developers.

Initially \emph{(1)} the custom programs should identify the topology of the virtual network and its VMs, by invoking \textit{list} APIs, and iterate over these resources to identify their mappings with the physical resources. In the case of the small network, it should identify 2 network interfaces at the guest level and 6 at the hypervisor level (i.e., 2 for virtual NICs, 2 for the router, 4 for the two subnets). This would require the invocation of 6 different APIs using about $\sim300$ \emph{LOC} of Python code. Then \emph{(2)}, the custom programs should generate an OpenStack Heat template to deploy VMs running the workload generators (e.g., iPerf). In our hypothetical scenario, the template would consist of $\sim 200$ \emph{LOC} in YAML. Alternatively, if OpenStack Heat is unavailable, the custom programs should create an image, a keypair, a security group, a floating IP address, and a block storage device or volume, by using 10 different APIs. Afterward \emph{(3)}, we estimate $\sim200$ \emph{Python LOC} of the custom programs for coordinating the execution and termination of workload generators, to prepare the VMs (e.g., to flush old data from guests), and to collect logs. Finally \emph{(4)}, to inject the simplest type of faults (e.g., network loss, delay, and corruption), the custom programs need to run external utilities (e.g., the {\lmttfont tc} command), by targeting the appropriate physical network resources that map to the targeted virtual resources. Moreover, the program should also run external utilities to disable the injections and restore normal connectivity. For more complex fault types (e.g., the virtual configuration faults in our solution, which drop resources from the virtual network), the custom programs would need to invoke APIs to modify and to restore the virtual network, which is difficult due to the dependencies between the virtual resources that need to be modified recursively (e.g., to drop a subnet, its virtual nodes also need to be removed and restored).

In total, the custom programs would result in $\sim500$ of Python LOC. It is important to note that these programs would be only applicable for the specific virtual network under test, and that would need to be changed to accommodate a different virtual network. Thus, the custom program can grow up to thousands of Python LOC to handle more complex scenarios, such as the two case studies presented in the experimental section. Our estimate is conservative, as we did not consider error handling, logging, and other desirable features (e.g., configurability) of the custom-made programs. Moreover, the estimate does not account for the efforts of configuring the physical nodes to enable the custom program to launch commands across the infrastructure and to collect data. Overall, the efforts needed to develop and test these custom programs can take several days of developers' time. 
Using \toolname{}, developers can fully automate all of these operations, by using a reduced set of REST APIs or the visual dashboard. \toolname{} consists of reusable fault injection components, which dynamically configure the fault injections according to the topology of the virtual networks. These components reuse APIs and template-based orchestration from the infrastructure, in a generic and configurable way. In the simplest scenarios, only 4 REST APIs are needed to start, wait for termination, stop, and save logs.

%

%
%
%
%
%



\vspace{3pt}
\noindent
$\rhd$ \textbf{Tenant isolation.} \toolname{} can inject faults in multi-tenancy scenarios, without any impact on tenants beyond the one that conducts the experiments. 
\revision{Multi-tenancy is implemented in several Linux-based cloud management platforms (like OpenStack and Kubernetes) by leveraging namespaces provided by the Linux kernel. The purpose of each namespace is to wrap a particular global system resource in an abstraction that makes it appear to the processes within the namespace that they have their own isolated instance of the global resource. One of the overall goals of namespaces is to support the implementation of containers (thus Kubernetes), which includes a group of processes with the illusion that they are the only processes on the system.
The \toolname{} OpenStack implementation leverages both namespaces and the built-in mechanisms for handling multi-tenancy provided by OpenStack. In particular, OpenStack offers multi-tenancy through resource separation (compute, network and storage) via the so-called \textit{projects}, which allow users to share virtual resources while keeping logical separation. For example, network traffic separation is provided by creating different VLAN IDs for Neutron networks of different projects. Actually, network separation is translated to different network namespaces used at a lower level. ThorFI masks these aspects to support users in configuring and performing fault injection tests.}

We performed additional experiments to assess the isolation among tenants, which were not previously presented in detail for the sake of brevity. In these experiments, we deployed both a tenant to run the case study software of the previous sections, and additional tenants that generate synthetic traffic using the workload generators included in \toolname{}. We did not observe any degradation of latency or throughput in the network traffic of the non-injected tenants, since \toolname{} correctly injects faults only in the physical network resources mapped to the targeted tenant. Moreover, in our tests \toolname{} was always able to restore the normal state of the virtual network after the injection, even in the case of complex faults that delete virtual resources from the network.

\vspace{3pt}
\noindent
$\rhd$ \textbf{Portability.} As mentioned in \sectionautorefname{}~\ref{sec:implementation}, \toolname{} adopts a generic architecture that can be ported to several virtualization technologies and cloud computing platforms. As a representative example, \toolname{} can be adapted to support the popular container orchestration platform Kubernetes \cite{kubernetes}, by mapping the virtual network to the resources used by the Kubernetes infrastructure. Kubernetes organizes the data center into clusters of worker nodes, which run containerized applications in \textit{Pods}. Each pod has its own IP address and includes one or more containers with shared storage and network resources. Kubernetes can leverage different implementations of networking, but imposes that pods on a node can communicate with all other pods on all other nodes. In particular, Kubernetes supports a large spectrum of networking options, ranging from classical ones (e.g., L2 networks and Linux bridging, OpenVSwitch, OVN), Kubernetes-specific ones (e.g., Kube-router, Kube-OVN, and k-vswitch) and commercial network backends (e.g., Calico by Tigera, Contrail by Juniper). Regardless of the specific networking approach, \toolname{} can be extended by modifying the APIs for scanning the network topology and injecting network faults against network interfaces underlying the target cloud applications. To identify the network resources to be targeted (i.e., the {\lmttfont thorfi\_item\_map} table, see \subsectionautorefname{}~\ref{subsec:thorfi_architecture}), the Kubernetes implementation would need to identify the network namespace IDs assigned both to the target Kubernetes pod and the node that hosts this pod. Each pod's network namespace communicates with the node's root network namespace through a virtual Ethernet pipe ({\lmttfont veth} device). Thus, once obtaining information about namespaces, \toolname{} can reuse common user-space tools provided by Linux (e.g., {\lmttfont nsenter} and {\lmttfont ip addr}) to correlate which {\lmttfont veth} device is paired with a particular pod. 
A similar approach can be applied for other virtualization technologies, such as the Xen and KVM hypervisors, to map guest virtual resources to network interfaces in the host.


\section{Conclusion}
\label{sec:conclusion}

In this work, we presented a novel solution for fault injection as-a-service in virtual networks. We designed this solution for cloud tenants, in order to support them at assessing the resiliency of their cloud applications, while avoiding side effects on the other tenants. The solution integrates with various cloud infrastructure management software. We presented the implementation of the solution in the context of the OpenStack platform and released it as open-source software.

We exposed two case studies to show use cases for the proposed fault injection solution. In the first case study, we used the solution to extend functional test cases for virtualized network functions, to evaluate the compliance of the system-under-test to protocol standards. These standards entail many corner cases, which can only be covered through fault injection. In our experiments on an NFV IMS configured to be highly available, we found that many test cases experience failures when faults are injected, pointing out flaws in its failure management mechanisms. In the second case study, we used the solution to collect performance measurements under faults, to support system designers in capacity and redundancy planning. We found that some faults are not well handled despite redundancy, and that the measurements enable quantitative evaluations of time-to-recovery.

\revision{Concerning the limitations of the proposed framework, ThorFI is designed to be used in a cloud management platform, which may not be available in all scenarios (e.g., in the case of systems running on bare metal or dedicated systems). Further, the cloud platform could not be accessible (e.g. public cloud providers that do not allow users to install tools such as ThorFI in their own infrastructures).}

Potential future extensions of the proposed solution are to make it support other cloud infrastructure management software and hypervisors, such as VMware vSphere and ESXi. Moreover, we foresee the use of this fault injection solution as a basis for new approaches to make virtualized network services more resilient. Following the DevOps paradigm, virtualized network functions are going to be released and update at a fast pace, to meet time-to-market goals and to accommodate for new, value-added features. Integrating fault injection in the cloud management software enables a quick and seamless reliability assessment of these network functions, thus enabling faster releases with higher confidence. Moreover, the feedback from fault injection could be automatically fed to failure management solutions, e.g., to train fault localization and to make it follow the updates in the virtual network, and to calibrate tunable parameters for failure detection.

\bibliography{bibliography}

\section*{}
\textbf{Domenico Cotroneo} (Ph.D.) is a Full professor at the University of Naples Federico II, Italy. His research interests include software fault injection, dependability assessment, and field-based measurement techniques.
\subsection*{}
\textbf{Luigi De Simone} (Ph.D.) is a Research Fellow at the University of Naples Federico II, Italy. His research interests include dependability benchmarking, fault injection testing, virtualization technologies reliability and its application on safety-critical systems. 
\subsection*{}
\textbf{Roberto Natella} (Ph.D.) is an Assistant Professor at the University of Naples Federico II, Italy. His research interests include dependability benchmarking, software fault injection, robustness and security testing, software aging and rejuvenation, and their application in OS and virtualization technologies.

\end{document}